\documentclass[preprint,prd,aps,floats,nofootinbib,showpacs,floatfix]{revtex4-1}

\usepackage{graphicx}
\usepackage[colorlinks=true, citecolor=blue, urlcolor = blue, linkcolor= blue, bookmarks=true]{hyperref}

\newcommand{\PRLett}[3]{Phys. Rev. Lett. {\bf #1}, #2 (#3).}
\newcommand{\PRD}[3]{Phys. Rev. D {\bf #1}, #2 (#3).}
\newcommand{\PRC}[3]{Phys. Rev. C {\bf #1}, #2 (#3).}
\newcommand{\PLB}[3]{Phys. Lett. B {\bf #1}, #2 (#3).}
\newcommand{\EPJA}[3]{Eur. Phys. J. A {\bf#1}, #2 (#3).}

\newcommand{\NPA}[3]{Nucl. Phys. A {\bf #1}, #2 (#3).}
\newcommand{\NPB}[3]{Nucl. Phys. B {\bf #1}, #2 (#3).}
\newcommand{\JPhysG}[3]{J. Phys. G {\bf #1}, #2 (#3).}

\newcommand{\gam }{\ensuremath{\gamma }}
\newcommand{\gs }{\ensuremath{\sigma }}     
\newcommand{\gz }{\ensuremath{\zeta }}       
\newcommand{\gd }{\ensuremath{\delta }}      
\newcommand{\go }{\ensuremath{\omega }}   
\newcommand{\gr }{\ensuremath{\rho }}         
\newcommand{\gl }{\ensuremath{\lambda}}     

\newcommand{\gn }{\ensuremath{\eta}}         
\newcommand{\gp }{\ensuremath{\phi }}         

\newcommand{\gX }{\ensuremath{\Xi }}          
\newcommand{\gS }{\ensuremath{\Sigma }}   
\newcommand{\gL }{\ensuremath{\Lambda }} 
\newcommand{\gG }{\ensuremath{\Gamma }}   

\newcommand{\fb }{\ensuremath{f_{B}}}

\newcommand{\llags}[1]{\mathcal{L}_{#1}} 

\newcommand{\dmucon}{\ensuremath{\partial^\mu}} 
\newcommand{\dmucov}{\ensuremath{\partial_\mu}} 

\newcommand{\bplus }{\ensuremath{B^+}}
\newcommand{\bminus }{\ensuremath{B^-}}
\newcommand{\bzero }{\ensuremath{B^0}}
\newcommand{\bbarzero }{\ensuremath{{\bar B}^0}}

\newcommand{\dm }{\ensuremath{D} meson}
\newcommand{\Bm }{\ensuremath{B} meson}
\newcommand{\Bbarm }{\ensuremath{\bar B} meson}
\newcommand{\BandBbarm }{\ensuremath{B} and \ensuremath{\bar B} meson}

\newcommand{\mb }{\ensuremath{m_B}}

\newcommand{\lag }{Lagrangian}
\newcommand{\lagd }{Lagrangian density}

\newcommand{\wtt }{Weinberg-Tomozawa term}

\newcommand{\grz }{\ensuremath{\rho_0}}
\newcommand{\gri }{\ensuremath{\rho_i}}
\newcommand{\grsi }{\ensuremath{\rho_i^s}}
\newcommand{\grb }{\ensuremath{\rho_B}}

\newcommand{\grsub}[1]{\ensuremath{\rho_{#1}}}
\newcommand{\grssub}[1]{\ensuremath{\rho_{#1}^s}}

\newcommand{\isoap }{isospin asymmetry parameter}
\newcommand{\dg}{\dagger}
\newcommand{\chigrp}{\ensuremath{SU(3)_L\times SU(3)_R}} 

\begin{document}

\title{\protect \boldmath 
Open bottom mesons in hot asymmetric hadronic medium 
}
\author{Divakar Pathak}
\email{dpdlin@gmail.com}
\affiliation{Department of Physics, Indian Institute of Technology, Delhi, 
Hauz Khas, New Delhi $-$ 110 016, India}

\author{Amruta Mishra}
\email{amruta@physics.iitd.ac.in}
\affiliation{Department of Physics, Indian Institute of Technology, Delhi,
Hauz Khas, New Delhi $-$ 110 016, India}

\begin{abstract}
The in-medium masses and optical potentials  of $B$  and ${\bar B}$ mesons 
are studied in an isospin asymmetric, strange, hot and dense hadronic 
environment using a chiral effective model. The 
chiral $SU(3)$ model
originally designed for the light quark sector, is generalized to 
include the heavy quark sector 
($c$ and $b$) to derive the interactions 
of the $B$ and $\bar B$ mesons with the light hadrons. Due to large 
mass of bottom quark, we use only the empirical form of these 
interactions for the desired purpose, 
while treating the bottom degrees of freedom to be frozen in the medium. 
Hence, all medium effects are due to the in-medium interaction of the 
light quark content of these open-bottom mesons. Both \BandBbarm s 
are found to experience net attractive interactions in the medium, 
leading to lowering of their masses in the medium. 
The mass degeneracy of particles and antiparticles, 
(\bplus,\ \bminus) as well as (\bzero,\ \bbarzero), is observed to be
broken in the medium, due to equal and opposite contributions from a 
vectorial Weinberg-Tomozawa interaction term. Addition of hyperons 
to the medium lowers further the in-medium mass for each of these 
four mesons, while a non-zero isospin asymmetry is observed 
to break the approximate mass degeneracy of each pair of isospin 
doublets. These medium effects are found to be strongly density 
dependent, and bear a considerably weaker temperature dependence. 
The results obtained in the present investigation are compared 
to predictions from the quark-meson coupling model, 
heavy meson effective theory, and the QCD Sum Rule approach.
\end{abstract}

\pacs{12.39.Fe; 12.38.Lg; 11.30.Rd; 21.65.Jk; 21.65.Cd}

\maketitle

\begin{section}{Introduction}\label{introduction}

It is widely recognized that the properties of hadrons in the medium are 
different from their behavior in vacuum 
\cite{Rev_1997,RMP2010,Li_Rev_99,glen}. 
The low-energy dynamics of QCD (i.e. of the hadronic phase) 
is governed principally by chiral symmetry, whose spontaneous breaking 
leads to a non-vanishing scalar condensate in vacuum. 
The properties of the hadrons containing light quark(s)
depend on the light quark condensate and 
are modified in the medium in accord with it \cite{Rev_1997}. 
While addressing all these questions of hadronic in-medium 
behavior, one is essentially in the non-perturbative regime of QCD.
In this regime, the perturbative techniques no longer are applicable 
and there exist diverse techniques to study the medium modifications of the 
hadrons, which can be broadly grouped into: the coupled channel approach 
\cite{Oset_PRL98,Oset_NPA98,mam_cpldchnl_2004,hofmannlutz}, QCD sum rules
\cite{QSR_original,hayashigaki,arvQSR,Hilger_QSR_D_Bm},
 quark-meson coupling model \cite{QMC1,QMC2,Bm_QMC_Tsushima}, 
relativistic mean field approaches based on the Walecka model 
\cite{serotwalecka} and its subsequent extensions, and, 
the method of chiral-invariant \lag s. This multitude of approaches 
provides the additional advantage that while questions of validity 
of theoretical approaches are conclusively settled only by experimentation, 
in the temporary absence of experimentation, a comparison between 
two independent well-founded approaches is still a healthy way 
of ascertaining whether or not one is on the right track. 
These medium effects have been predicted to have several 
important consequences, which also reflects the potential 
significance of this problem. These range from antikaon 
condensation \cite{kaplan_nelson}, sub-threshold production 
of particles, overall enhancement in dilepton production 
\cite{li_ko_brown_prl95,li_ko_brown,li_ko}, extra decay channels 
and consequent suppression in the yield of the parent particle 
(e.g. $J/\Psi$ suppression \cite{Jpsi_suppression_decays_mesons}, 
for which this is a possible mechanism), unequal particle ratios 
for isospin pairs in heavy ion collision experiments 
\cite{ParticleRatiosJPhysG_1}, as well as production asymmetry
\cite{Braun_Munzinger_Production_Asymmetry} for antiparticles. 

One approach that has been vigorously pursued in 
the past few years 
is to treat these medium effects from the point of view of 
a phenomenological, effective, hadronic \lag \ based 
on the QCD symmetries (in particular, the chiral symmetry)
and symmetry-breaking patterns \cite{Pap_prc99, Zsch}. 
While this was originally devised as a further step 
in the evolution of this effective \lag \ approach, 
which explicitly accounted for these features 
while these were not a part of the earlier hadrodynamical 
(Walecka-type) models, this has grown into an extremely 
productive method which has been fruitfully applied 
to understand the behavior of matter under extreme conditions 
of density and temperature. In its original incarnation, 
this model was used to successfully describe nuclear matter, 
finite nuclei, neutron stars and hypernuclei \cite{Zsch}. 
Subsequently, this was used to understand the in-medium behavior 
of vector mesons \cite{mamZsch_vector2004, mam_balazs_vector2004}, 
and more extensively, that of kaons and antikaons 
\cite{mamK2004, mam_kaons2006, mam_kaons2008, sambuddha1, sambuddha2}, 
which is natural, given the fact that this model was specifically 
tailored for the chiral $SU(3)$ situation.
Of late, this approach has been extended to the charm sector 
in pseudoscalar mesons as well, by generalizing this effective 
\chigrp \ model to $SU(4)$, and applied to 
study the in-medium behavior of \dm s
 \cite{mamD2004, arindam, arvDprc, arvDepja}.
 However, since $SU(4)$ symmetry is badly broken owing
 to the large mass of the charm quark, these analyses
 only used this $SU(4)$ symmetry to derive the empirical
 form of the interactions, while the charm degrees of freedom
 were treated as frozen in the medium.    
Tacitly, therefore, each of these studies treats
a \dm \ as a heavy-light system of quarks and antiquarks, 
with the dynamics of the heavy quark frozen.
Such a system gets modified in a hadronic medium due to
 the interactions of the light $u$ and $d$ quarks (and anti-quarks)
of the open charm meson
 with the particles constituting the medium, and not because
 of the heavy quark content. However, from a physical perspective,
 if the heavy quark is to be treated as frozen, 
a \emph{light antiquark-bottom} pseudo-scalar meson (e.g. ${\bar u} b$)
 is similar to a \emph{light antiquark-charm} pseudo-scalar meson 
(e.g. ${\bar u} c$).
In the present investigation, we generalize the chiral effective
 approach to the bottom sector, and derive the interactions
of the $B$ and $\bar B$ mesons with the light hadrons
to determine the in-medium behavior of these \emph{light quark-bottom} 
meson systems. 
On the other hand, the heavy quarkonium systems, e.g., charmonium
and bottomonium states, due to the absence of any light quark
constituents, are modified in the medium due to their
interactions with the gluon condensates \cite{ arvDepja,leeko,leeko2}. 
A study of the
mass modification of the charmonium system in the medium 
arising from the medium modifications of the gluon condensate
has been recently generalized to study the bottomonium  states
in the medium \cite{DP_Bmonia}.

The topic of the properties of the $B(\bar B)$ mesons can also be
important 
in the study of  
\Bm \ diffusion, 
drag and propagation in a hot and/or dense hadronic matter. 
Heavy flavored mesons are considered to be valuable probes 
for analyzing the behavior of matter in hot and dense medium, 
in both the quark-gluon plasma and hadronic phase \cite{Tolos_Bm}. 
This is because they carry a heavy (charm or bottom) quark which 
has a special significance as regards the experimental characterization 
of matter formed in a high energy collision. Based on explicit solutions 
from a Langevin model formulated to study the questions of transport
 and thermalization of heavy quarks in a quark-gluon plasma 
\cite{MooreTeaney_HeavyQuarkThermalization}, it has been reasoned 
that the thermal relaxation time for the heavy quarks is significantly
larger than that of lighter quarks, due to which these heavy quarks 
are likely not to reach an equilibrium with their ambience, and hence, 
(upon subsequent hadronization) still retain information about the 
initial stages of the heavy ion collision, when these were produced
\cite{Fries_heavyQ_PRC2012}. Thus, analyzing these heavy flavored 
mesons is an indirect and efficient 
way of finding out about the early stages of these collisions.
This realization has led to a flurry of recent 
activity \cite{Fries_heavyQ_PRC2012, JaneAlam_Bm,
Tolos_Bm, Bm_Transport_MesonGas,HeFriesPLB_May2014}, 
studying both, the utility of open-bottom mesons as a probe, 
as well as their transport properties in a hadronic medium. 
Especially, Refs. \cite{Tolos_Bm,Bm_Transport_MesonGas} establish
that not just that the open-bottom mesons do not thermalize 
at the the kind of energies one encounters at LHC and RHIC,
 \Bm s are more suited to serve as probes as compared to
 the charmed mesons, in heavy-ion collision experiments
 (from the considerations of relaxation length). 
This may be perceived as a good news, considering the kind
 of impetus the first generation $B-$factory experiments has
 provided to $b$-physics. In recent years, both the BaBar
 experiment \cite{BaBar_Offical_Webpage} at the PEP-II
 $e^+ e^-$ energy collider in SLAC, and the BELLE
 experiment \cite{BELLE_Official_Webpage, BELLE_PTEP2012}
 at the KEKB $e^+ e^-$ energy collider, have utilized
 these respective high-luminosity experimental facilities to
 significantly improve the current understanding of bottom-flavored
 hadrons. The next generation $B-$factory experiments
 (BELLE-II \cite{BELLE2_Official_Webpage}, currently in the pipeline)
 are expected to enhance the experimental situation still further,
 considering a $40-$fold increase in the instantaneous luminosity
 proposed in the SuperKEKB upgrade project of KEK \cite{BELLE2_DESY,
 BELLE2_Technical_Design_Report}. 
The observation of the hadrons with the heavy b-quark/antiquark
has initiated studies of their in-medium behavior.
 One can easily reason on the basis of the strong density dependent
 medium effects already observed for the particles having light 
quark content, that corresponding medium effects for the \BandBbarm s 
would also be substantial, due to similar light quark content.
 Therefore, for a full appreciation of the behavior of the \BandBbarm s 
in such conditions, one must consider the effect of medium modifications
 of these mesons under such conditions. However, 
systematic treatments of such medium effects for these mesons
 have barely started pouring in, and there is need for more work
on this subject.
Apart from the works mentioned above, devoted specifically
 to the issue of their transport properties in the medium,
 there exist analyses of \Bm \ in-medium behavior using
 the QCD Sum Rule method \cite{Hilger_QSR_D_Bm},
 and also using the quark meson coupling model
 \cite{Bm_QMC_Tsushima}. Yasui and Sudoh have recently contributed
 considerably to this field, by analyzing the \Bm \ properties
 within three different approaches $-$ 
within heavy meson effective theory with $1/M$ corrections
 \cite{YasuiSudoh_Bm_2014}, by considering an effective
 \lag \ for $B-N$ interaction due to pion exchange
 \cite{YasuiSudoh_Bm_2013_1}, and especially the analysis
 of Ref.\cite{YasuiSudoh_Bm_2013_2}, where these were treated
 as heavy impurities embedded in a finite density medium,
 based on symmetry considerations. This third work
 \cite{YasuiSudoh_Bm_2013_2}, in particular,
 offers a unique perspective on this issue, as 
this physical situation is likened to the famous `Kondo problem'
 in condensed matter physics. In the present work, we study the
properties of the $B(\bar B)$ mesons in isospin asymmetric 
hyperonic matter at finite temperatures, arising from their
interactions with the light hadrons in a chiral effective model.   

We organize this article as follows: in section II, we outline 
the chiral \chigrp \ model, and its subsequent generalization, 
used in this investigation. In section III, the \lagd \ for the 
\BandBbarm s, within this model, is explicitly written down 
and is used to derive their dispersion relations in the medium. 
In section IV, we describe and discuss what the preceding 
formulation implies for the in-medium properties of \BandBbarm s 
and mention the possible implications of these medium effects. 
Finally, we summarize the findings of the present investigation 
in section V. 
\end{section}

\begin{section}{The Chiral effective model}

The current investigation is based on a generalization of the chiral 
\chigrp \ model \cite{Pap_prc99} designed for the light hadrons,
to the heavy quark (charm and bottom) sector. A detailed exposition 
of the model can be found 
in Refs. \cite{Pap_prc99,Zsch}, but its main features are summarized here, 
for conciseness. 
This is a relativistic field theoretical model of interacting baryons
and mesons, wherein the form of the interactions 
is dictated by chiral invariance. 
In this treatment, a nonlinear realization of chiral symmetry is adopted, 
which is in line with the approach successfully followed by Weinberg 
\cite{weinberg67,weinberg68} for the $SU(2)_L \times SU(2)_R$ case. 
The same was generalized to arbitrary compact Lie groups and a general
 formulation for the construction of chiral-invariant \lag s was
 given in Refs. \cite{coleman1,coleman2,bardeenlee}. 
Also, the scale symmetry (invariance) which is broken in QCD, 
is introduced in the chiral model through a scalar dilaton field, 
$\chi$ \cite{brownrho_PRL1991,ellis,schechter}. 
The expectation value of the dilaton field gets related
to the expectation value of the scalar gluon condensate, 
as can be seen via a comparison of the trace of the energy-momentum 
tensor for the QCD case and for the chiral effective model
\cite{brownrho_PRL1991,ellis,arindam,arvDprc,arvDepja}. 
It may be noted here that the scalar gluon condensate as calculated 
through the chiral effective model \cite{arindam,arvDprc,arvDepja} 
was used to calculate the mass shifts of the charmonium states 
through QCD second order Stark effect \cite{arvDepja}. 
The effect of the twist 2 gluon condensates, as obtained
from the medium change of the dilaton field calculated 
within the chiral effective model, on the in-medium masses 
of the $J/\psi$ and $\eta_c$, has also been calculated 
using QCD sum rule approach \cite{arvQSR}.
The results obtained for the mass shifts of the charmonium states
from the gluon condensates obtained within the chiral effective model,
at low densities, were observed to be similar to as obtained
using the gluon condensates of linear density approximation 
\cite{leeko,klinglkimlee}.

The general expression for the \lagd \ in this chiral effective model
 has the following form: 
\begin{equation}
{\cal L} = {\cal L}_{\rm kin}+\sum_{W} {\cal L}_{\rm BW} + {\cal L}_{\rm vec} 
+ {\cal L}_{0} + {\cal L}_{\rm scale \; break}+ {\cal L}_{\rm SB}
\label{genlag_model}
\end{equation}
In eqn.(\ref{genlag_model}), $\llags{\rm kin}$ is the kinetic energy term.
 $\llags{\rm BW}$ is the baryon-meson interaction term, where the index $W$
 covers both spin$-0$ (scalar) and spin$-1$ (vector) mesons. Here,
 the baryon masses are generated dynamically, through the
 baryon-scalar meson interactions. $\llags{\rm vec}$ concerns
 the dynamical mass generation of the vector mesons through couplings
 with scalar mesons, apart from bearing the self-interaction terms
 of these mesons. 
$\llags{0}$ contains the meson-meson interaction terms,
and ${\cal L}_{\rm scale  \; break}$ incorporates the
scale invariance breaking of QCD through a logarithmic potential. 
Finally, the explicit symmetry breaking of $U(1)_A$, $SU(3)_V$
 and chiral symmetry is incorporated in this effective hadronic
 model through the term $\llags{\rm SB}$.

An analysis of the medium modifications of pseudoscalar mesons
due to their interactions with the baryons (nucleons and hyperons)
and scalar mesons, requires the assessment of the following 
contributions to the Lagrangian density -
\begin{equation}
{\cal L}_{\rm pseudoscalar}= 
\llags{\rm WT} + \llags{\rm SME} + \llags{\rm 1^{\rm st} Range} 
+ \llags{\rm d_1} + \llags{\rm d_2},
\label{genlag}
\end{equation}
where, the first term is the vectorial Weinberg-Tomozawa term,
the second term arises due to the scalar meson exchange,
and the last three terms are the range terms. The Weinberg-Tomozawa
term corresponds to the leading order contribution, and, the scalar 
exchange term and range terms correspond to the the next to 
leading order contribution in the chiral perturbation expansion
\cite{mam_kaons2006,mam_kaons2008,arindam,arvDprc,arvDepja}.
In the above, the \wtt, $\llags{\rm WT}$ is given as - 
\begin{eqnarray}
\llags{\rm WT} & = & -\frac{1}{2}\Big[ \bar{B}_{ijk} \,\gam^\mu\,
\left( {({\gG_{\mu}})_l}^k\,B^{ijl}+ 
2\ {({\gG_{\mu}})_l}^j\,B^{ilk}\right) \Big],
\label{LWTexpression}
\end{eqnarray}
where repeated indices are summed over. The same originates from the 
kinetic energy term ($\llags{\rm kin}$ in Eq. (\ref{genlag_model})) 
in the chiral model. The tensor $B^{ijk}$,
which is antisymmetric in the first two indices, represents 
the baryons \cite{hofmannlutz}. The indices $i, \ j$ and $k$ 
run from 1 to 5, and one can read off the quark/antiquark content 
of a baryon state, $B^{ijk}$ as well as of the pesudoscalar mesons
given as the matrix elements of the pesudoscalar matrix,
$M$ occurring in the expression $\Gamma_\mu$ as given in the 
following, with the identification: 
$1 \leftrightarrow u, 2 \leftrightarrow d, 
3 \leftrightarrow s, 4 \leftrightarrow c,
5 \leftrightarrow b$. 
However, in the current investigation, 
just like the charmed baryons \cite{arvDepja}, 
the medium modifications of the heavier (bottomed) baryons
have not been accounted for, to study the in-medium
properties of the $B$ and $\bar B$ mesons.
In equation (\ref{LWTexpression}), 
$\gG_\mu$ is defined as  
\begin{equation}
\gG_\mu \ = -\frac{i}{4}\left [\Big( u^{\dg} \big( \dmucov u \big)
 - \big( \dmucov u^{\dg} \big) u\Big ) 
+ \Big ( u \big( \dmucov u^{\dg} \big)
 - \big( \dmucov u \big) u^{\dg} \Big) \right],
\label{Gammamu1}
\end{equation}
where the unitary transformation operator, $u$, is given as - 
\begin{equation}
u = \exp \left( \frac{iM}{\sqrt{2} \gs_0} \gamma_5 \right),
\end{equation}
where $M$ represents the matrix of pseudoscalar mesons, constructed 
as $M = (M^a \gl_a / \sqrt{2})$, where $M^a$ represents the field 
corresponding to $a^{th}$ pseudoscalar meson, and the $\gl_a$'s 
refer to the generalized Gell-Mann matrices.
$\llags{\rm SME}$ is the scalar meson exchange term, which is obtained 
from the explicit symmetry breaking term ($\llags{\rm SB}$ 
in Eq. (\ref{genlag_model})) - 
\begin{equation}
\llags{\rm SB}  =  -\frac{1}{2} \ {\rm Tr} 
\left( A_p \left(uXu + u^{\dg}Xu^{\dg}\right) \right),
\label{LSBexpression}
\end{equation}
where, 
\begin{eqnarray}
A_p & = & \frac{1}{\sqrt{2}} \ {\rm diag} 
\Big[ m_{\pi}^2 f_{\pi}, m_{\pi}^2 f_\pi, 
\left(2 m_K^2 f_K -m_{\pi}^2 f_\pi \right), 
\left(2 m_D^2 f_D - m_{\pi}^2 f_\pi \right), 
\left(2 m_B^2 f_B - m_{\pi}^2 f_\pi \right) \Big].
\end{eqnarray} 
The constants for the above expression are chosen so that, 
in conjunction with the fitted vacuum expectation values 
of the scalar fields, the  PCAC relations are respected. 
The remaining terms in eqn.(\ref{genlag}) are the range terms, which 
have the basic structure $(\dmucov M) (\dmucon M)$.
The first range term is obtained from the kinetic energy term ($\llags{\rm kin}$ in Eq.(\ref{genlag_model})) of the pseudoscalar mesons in the chiral model \cite{Pap_prc99}, and goes as :
\begin{equation}
\llags{\rm 1^{\rm st} Range} =  {\rm Tr} \big(u_{\mu} X u^{\mu}X 
+X u_{\mu} u^{\mu} X \big) 
\label{Lfirstrangeexpression}
\end{equation}
where $u_{\mu}$ is defined in terms of the unitary transformation 
operator $u$, and its derivatives, as:
\begin{equation}
u_{\mu}  = -\frac{i}{4}\left [\Big( u^{\dg} \big( \dmucov u \big)
 - \big( \dmucov u^{\dg} \big) u\Big ) 
- \Big ( u \big( \dmucov u^{\dg} \big)
 - \big( \dmucov u \big) u^{\dg} \Big) \right ],
\end{equation} 
The other two range terms in Eq.(\ref{genlag}) are the $d_1$ and $d_2$ range terms, 
whose expressions are given below.
\begin{equation}
\llags{\rm d_1} = \frac{d_{1}}{4}\Big( \bar B_{ijk} B^{ijk}
{{(u_\mu)}_{l}}^{m}{{(u^\mu)}_{m}}^{l}\Big)
\label{Ld1expression}
\end{equation}
\begin{equation}
\llags{\rm d_2} =\frac{d_{2}}{2}\Big[ \bar B_{ijk} {{(u_\mu)}_{l}}^{m} 
\ \Big({{(u^\mu)}_{m}}^{k}B^{ijl} + 2{{(u^\mu)}_{m}}^{j}B^{ilk}\Big)\Big] 
\label{Ld2expression}
\end{equation}
(repeated indices summed, as before). 
The $d_1$ and $d_2$ terms are the range terms which have been constructed 
from the baryon and pseudoscalar meson octets, within the chiral
SU(3) model, to study the in-medium properties of the kaons and antikaons 
\cite{mam_kaons2008}. These terms were then generalized to SU(4)
to study the D-mesons \cite{arindam, arvDprc,arvDepja} and were written
in the above form using the tensorial motations for the baryons as well
as pseudoscalar mesons, since the baryons belong to a 20-plet and the
mesons belong to 15-plet. In the present work, the interactions
for the B mesons have been written down in a similar manner,
including also the b-quarks. 
We make use of the mean field approximation 
\cite{serotwalecka, Zsch} to study hadron properties at finite densities
and temperatures. 
Thus, we approximate for every scalar field $\phi$ and vector field $V^{\mu}$, 
\begin{eqnarray}
\phi \rightarrow \left<\phi \right> \equiv \phi_0, \ \ \ V^{\mu}
 (\equiv (V_0, {\vec V}) )\rightarrow \left \langle V^{\mu} \right
\rangle  \equiv (V_0, 0), 
\label{MFA}
\end{eqnarray}
where $\phi_0$ and $V_0$ are constants independent of space and time. 
$X$, occurring in 
eqns.(\ref{LSBexpression}) and (\ref{Lfirstrangeexpression}), 
is the scalar meson multiplet, which in the mean field approximation,
is given as,
\begin{equation}
X \ = \  {\rm diag} \left[ \frac{(\gs + \gd)}{\sqrt{2}},
 ~\frac{(\gs - \gd)}{\sqrt{2}}, ~\gz, ~\gz_{c}, ~\gz_{b} \right]
\end{equation}
In the above, $\gs (\sim ({\bar u}u + {\bar d}d)), \ \gz (\sim {\bar s}s), \ \gz_c (\sim {\bar c}c)$, and $\gz_b (\sim {\bar b}b)$ are the non-strange, strange, charmed and bottomed scalar-isoscalar mesons, and $\gd (\sim ({\bar u}u - {\bar d}d))$ is the non-strange scalar-isovector meson. Within the mean field approximation, the equations of motion for 
the scalar and vector mesons, are derived, which are 
subsequently used in this investigation. 
It has been realized over a period of time, that this approximation, 
which is a vast simplification over the general case, is sufficient
for a reasonable description of hadronic in-medium properties 
\cite{Zsch, mamD2004, mam_kaons2006, mam_kaons2008, sambuddha1, 
sambuddha2, arvDprc, arvDepja, arindam}.  
We then write down the explicit expression
for the Lagrangian density describing the interaction 
of the $B$ and $\bar B$ mesons with the light 
hadrons and the in-medium dispersion relations 
of the \BandBbarm s obtained from this interaction Lagrangian density,
in the next section. 
\end{section}

\begin{section}{\protect \boldmath \texorpdfstring{$B$}{B} and
 \texorpdfstring{${\bar B}$}{Bbar} Mesons in Hadronic Matter}
The \lagd \ for the \BandBbarm s in an isospin-asymmetric, strange,
 hadronic medium reads -
\begin{equation}
{\cal L}^B_{\rm total} = {\cal L}^B_{\rm free} + {\cal L}^B_{\rm int},
\end{equation}
where this $\llags{\rm free}^B$ is simply the free \lagd \ for the two pairs
 of complex scalar fields corresponding to the (\bplus,\ \bminus) and
 (\bzero,\ \bbarzero) mesons:  
\begin{eqnarray}
\llags{\rm free}^B & = &  \left( \dmucon \bplus\right) \left(\dmucov \bminus\right)
 -\mb^2 \left( \bplus \bminus \right) 
+ \left( \dmucon \bzero \right) \left(\dmucov \bbarzero \right)
 -\mb^2 \left( \bzero \bbarzero \right).
\end{eqnarray}
This $\llags{\rm free}^B$ can be recovered from the chiral model 
\lagd \ in vacuum,
from the expressions given by eqns.(\ref{LSBexpression}) 
and (\ref{Lfirstrangeexpression}),
by replacing $X$ by its vacuum expectation value, $X_0$, 
as
\begin{eqnarray} 
&&\left(\llags{\rm 1^{\rm st} Range}^B\right)_0  
=\left( \dmucon \bplus\right) \left(\dmucov \bminus\right) 
+ \left( \dmucon \bzero \right)
 \left(\dmucov \bbarzero \right),
 \nonumber\\ 
&&\left(\llags{\rm SME}^B\right)_0 \hspace{0.5cm} 
=-\mb^2 \left( \bplus \bminus + \bzero \bbarzero \right). 
\label{L_vacuum}
\end{eqnarray}
While the free \lagd \ for the \Bm s is borne out of these
two terms in vacuum, 
\begin{equation}
\llags{\rm free}^B = \left(\llags{\rm 1^{\rm st} Range}^B\right)_0 
+ \left(\llags{\rm SME}^B\right)_0 \ ,
\end{equation}
the finite density part of these two terms, given by eqns.(\ref{LSBexpression}) and
 (\ref{Lfirstrangeexpression}), contribute to the interaction \lagd.

The interaction \lagd \ for the $B$ and ${\bar B}$ mesons, within this 
generalized chiral effective approach reads:
\begin{equation}
{\cal L} ^{B}_{\rm int}  =
{\cal L}^B_{\rm WT} +{\cal L}^B_{\rm SME}+{\cal L}^B_{\rm range},
\label{L_Bmesons}  
\end{equation}
where,
\begin{eqnarray}
{\cal L}^B_{WT} &=&
\frac{-i}{8\fb^2} \Big [3\Big ({\bar p} \gamma^{\mu} p +{\bar n} \gamma^{\mu} n \Big) 
\Big( \big( (\dmucov \bplus) \bminus - \bplus (\dmucov \bminus) \big) 
+ \big( (\dmucov \bzero) \bbarzero - \bzero (\dmucov \bbarzero) \big) \Big) \nonumber \\
&+ & 
\Big( {\bar p} \gamma^{\mu} p -{\bar n} \gamma^{\mu} n \Big) 
\Big( \big( (\dmucov \bplus) \bminus - \bplus (\dmucov \bminus) \big)
- \big( (\dmucov \bzero) \bbarzero - \bzero (\dmucov \bbarzero) \big) \Big) \nonumber\\
&+ & 2 \Big( \bar{\Lambda} \gamma^{\mu} \Lambda + \bar{\Sigma}^{0} 
\gamma^{\mu} \Sigma^{0}\Big)  \Big( \big( (\dmucov \bplus) \bminus 
- \bplus (\dmucov \bminus) \big) + \big( (\dmucov \bzero) \bbarzero 
- \bzero (\dmucov \bbarzero) \big) \Big) 
\nonumber\\ &+ &  
 2 \ \Big(\bar{\Sigma}^{+} \gamma^{\mu} \Sigma^{+}
 + \bar{\Sigma}^{-} \gamma^{\mu} \Sigma^{-}\Big)
 \Big( \big( (\dmucov \bplus) \bminus 
- \bplus (\dmucov \bminus) \big) + \big( (\dmucov \bzero) \bbarzero 
- \bzero (\dmucov \bbarzero) \big) \Big) 
\nonumber\\ &+&  
 2 \ \Big(\bar{\Sigma}^{+}\gamma^{\mu} \Sigma^{+}
 - \bar{\Sigma}^{-}\gamma^{\mu} \Sigma^{-}\Big)
\Big( \big( (\dmucov \bplus) \bminus 
- \bplus (\dmucov \bminus) \big) - \big( (\dmucov \bzero) \bbarzero 
- \bzero (\dmucov \bbarzero) \big) \Big) 
\nonumber\\ &+&  
 \Big(\bar{\Xi}^{0}\gamma^{\mu} \Xi^{0}
 + \bar{\Xi}^{-}\gamma^{\mu} \Xi^{-}\Big) \Big( \big( (\dmucov \bplus) 
\bminus 
- \bplus (\dmucov \bminus) \big) + \big( (\dmucov \bzero) \bbarzero - \bzero (\dmucov \bbarzero) \big) \Big) \nonumber\\
&+& \Big(\bar{\Xi}^{0}\gamma^{\mu} \Xi^{0}
 - \bar{\Xi}^{-}\gamma^{\mu} \Xi^{-}\Big) \Big( \big( (\dmucov \bplus) 
\bminus 
- \bplus (\dmucov \bminus) \big) 
- \big( (\dmucov \bzero) \bbarzero - \bzero (\dmucov \bbarzero) 
\big) \Big)\Big],
\label{L_BmesonsWT}  
\end{eqnarray}

\begin{eqnarray}
{\cal L}^B_{\rm SME} &=&
 \frac{\mb^2}{2 \fb} \Big[\ (\gs' + \sqrt{2} \gz_b') 
\left( \bplus \bminus + \bzero \bbarzero \right) 
+ \gd \left( \bplus \bminus 
- \bzero \bbarzero \right) \Big]
\label{L_BmesonsSME}  
\end{eqnarray}
\begin{eqnarray}
{\cal L}^B_{\rm range}&=&
\left( \frac{-1}{\fb} \right) \Big[ \ (\gs' + \sqrt{2}\gz_b') 
\ \Big(\left( \dmucon \bplus \right) \left(\dmucov \bminus\right) 
+ \left( \dmucon \bzero \right) \left(\dmucov \bbarzero\right) \Big) 
\nonumber\\ 
& & 
\hspace{0.5cm} + 
\gd \Big( \left( \dmucon \bplus \right) \left(\dmucov \bminus\right) 
- \left( \dmucon \bzero \right) \left(\dmucov \bbarzero\right)\Big) \Big] 
\nonumber\\ &+ &
\frac {d_1}{2 f_B^2} \Big[ \big(\bar p p +\bar n n 
+\bar{\Lambda}\Lambda+\bar{\Sigma}^{+}\Sigma^{+}+\bar{\Sigma}^{0}
\Sigma^{0} 
+\bar{\Sigma}^{-}\Sigma^{-} +\bar{\Xi}^{0}\Xi^{0}
+\bar{\Xi}^{-}\Xi^{-}\big)
\nonumber\\ 
& & \hspace{1.5cm} 
\times  \Big ( (\partial _\mu {B^+})(\partial ^\mu {B^-}) 
+ (\partial _\mu {B^0})(\partial ^\mu {{\bar B}^0}) \Big)\Big] 
\nonumber\\
&+ &\frac {d_2}{4\fb^2} \ \Big[  \Big( 3( {\bar p} p 
+{\bar n} n )+ 2 \big( \bar{\Lambda}\Lambda
+ \bar{\Sigma}^{0}\Sigma^{0}\big )  
+2 \big((\bar{\Sigma}^{+}\Sigma^{+} 
+ \bar{\Sigma}^{-}\Sigma^{-})\big) 
+(\bar{\Xi}^{0}\Xi^{0}+ \bar{\Xi}^{-}\Xi^{-})\Big )
\nonumber \\ && \hspace{1.5cm} \times
\Big( (\dmucov \bminus)(\dmucon \bplus) 
+ (\dmucov \bbarzero)(\dmucon \bzero) \Big)
\nonumber \\  && \hspace{1cm}
+ \Big(\big ( {\bar p} p -{\bar n} n \big)
+2\big (\bar{\Sigma}^{+}\Sigma^{+} 
- \bar{\Sigma}^{-}\Sigma^{-}\big) 
+\big (\bar{\Xi}^{0}\Xi^{0} - \bar{\Xi}^{-}\Xi^{-}\big)\Big)
\nonumber \\ & & \hspace{1.5cm} \times
\Big( (\dmucov \bminus)(\dmucon \bplus) 
- (\dmucov \bbarzero)(\dmucon \bzero) \Big ) \Big ]
%
\label{L_Bmesonsrange}  
\end{eqnarray}
In Eq. (\ref{L_Bmesons}), the first term (with coefficient $(-i/8\fb^2)$
as given by equation (\ref{L_BmesonsWT})) is the \wtt, obtained 
from eqn.(\ref{LWTexpression}), the second term 
(with coefficient $(\mb^2/2\fb)$ as given by equation (\ref{L_BmesonsSME})) 
is the scalar meson exchange term, obtained from the explicit symmetry 
breaking term of the \lag \ (eqn.(\ref{LSBexpression})), 
third term is the range term given by equation (\ref{L_Bmesonsrange}).
The first range term (with coefficient $-1/\fb$) 
is obtained from eqn.(\ref{Lfirstrangeexpression}), and the other two
range terms (with coefficients $(d_1/2\fb^2)$ and $(d_2/4\fb^2)$, 
respectively) are the $d_1$ and $d_2$ terms, calculated from 
eqns.(\ref{Ld1expression}) and (\ref{Ld2expression}), respectively. 
Also, $\sigma^{\prime}\left( = \sigma - \sigma_{0}\right) $, 
$\zeta_{b}^{\prime}\left( = \zeta_{b} - \zeta_{b0}\right)$, 
and $\delta^{\prime}\left(  = \delta -\delta_{0}\right) $ are 
the fluctuations of the respective scalar fields, from their 
vacuum expectation values.  In writing down this form of the \lag, 
we have left out all the cross (bilinear) terms (for example, 
${\bar p}\gam^{\mu} n$, ${\bar n}\gam^{\mu}\gS^0$ etc.), 
since these do not contribute in the mean field limit. 
Additionally, from the transformation properties of Dirac bilinears, 
we recall that $\grssub{\ } = {\bar \psi} \psi$ is a scalar density, 
while the number density would be $\grsub{\ } = {\psi}^{\dg} \psi 
= {\bar \psi} \gam^0 \psi$, which is the zeroth component of the 
vector ${\bar \psi}\gam^{\mu} \psi$. 
As we had mentioned earlier, in eqn.(\ref{MFA}), in the mean field 
approximation, only the zeroth components of the vector fields 
contribute, that also as constants 
in space and time. Therefore, clubbing together the above arguments 
with eqn.(\ref{MFA}), mean field approximation for the baryons 
in this context gives:
\begin{eqnarray}
{\bar B_i} B_j \rightarrow & \left<{\bar B_i} B_j\right> & \equiv \gd_{ij} \grssub{i} \label{MFA_baryons_1}\\
{\bar B_i}\gam^{\mu}B_j \rightarrow & \left<{\bar B_i}\gam^{\mu}B_j\right> 
& = \gd_{ij} \left(\gd^0_{\mu} ({\bar B_i}\gam^{\mu}B_j) \right) 
\equiv \gd_{ij} \grsub{i}
\label{MFA_baryons}
\end{eqnarray}
where the indices $i$ and $j$ cover the entire `baryon octet' ($p$, $n$, $\gL$, $\gS^{\pm,0}$ and  $\gX^{0,-}$). 
As mentioned previously as well, in the present investigation, 
 we do not consider the effects of the still heavier (charmed and bottomed) baryons 
on the in-medium properties of the $B$ and $\bar B$ mesons.
The scalar and number densities of the $i$-th baryon ($i$=$p$, $n$, $\gL$, $\gS^{\pm,0}$ and  $\gX^{0,-}$), 
occurring in eqns.(\ref{MFA_baryons_1}-\ref{MFA_baryons}), 
are given by the expressions:
\begin{eqnarray}
\gri = \int d^3 k~(n_i (k) - \bar{n}_i (k)), \;\;\;\;
\grsi = \int d^3 k~\frac{m^*_i}{E^*_i (k)}~\big(n_i (k)
 + \bar{n}_i (k) \big), 
\end{eqnarray}
where $n_i (k)$ and $\bar{n}_i (k)$ represent the particle and 
antiparticle distribution functions, given by - 
\begin{equation}
n_{i}(k) \equiv n_{i}\big(k, \mu_i^*, T\big)
= \frac{1}{\exp{\left( \frac{E^*_i - \mu_i^*}{T}\right)} \pm 1},\;\;\;\;
\bar {n}_{i}(k) \equiv \bar{n}_{i}\big(k, \mu_i^*, T\big)
= \frac{1}{\exp{\left( \frac{E^*_i + \mu_i^*}{T}\right) } \pm 1}.
\end{equation}
In the above equations, $m_i^*$ is the effective mass of the $i^{th}$ 
baryon, given as - 
\begin{equation}
m^*_i = -\left( g_{\gs i}\gs + g_{\gz i}\gz + g_{\gd i}\gd \right),
\label{mi_star}
\end{equation}
and $\mu_i^*$ refers to the 
effective chemical potential of the $i^{th}$ baryon, given by the expression
\begin{equation}
\mu_i^* = \mu_i - \left( g_{\gr i}\tau_3 \gr + g_{\go i}\go 
+ g_{\gp i}\gp \right).
\label{effchempot}
\end{equation}

With the \lagd \ obtained in the mean field approximation, we use the Euler-Lagrange equation to determine 
the equations of motion for the \BandBbarm s. It can be readily observed from the form of the 
\lagd, eqns.(\ref{L_vacuum}) and (\ref{L_Bmesons}), that the equations of motion for \BandBbarm s 
would come out to be linear. Therefore, by assuming plane wave solutions 
$( \sim \ e^{i(\vec{k}.\vec{r} - \go t)} )$, it is possible to
 `Fourier transform' 
these equations of motion to obtain the in-medium dispersion relations for these mesons. 
These dispersion relations have the general form:
\begin{equation}
-\go^{2} + \vec{k}^2 + \mb^2 - \Pi (\go,|\vec{k}| ) = 0
\label{dispersion}
\end{equation}
where, \mb \ is the vacuum mass of the respective \Bm \ and $\Pi (\go,| \vec{k} |)$ is its 
\emph{self-energy} in the medium, the latter representing the contribution of medium effects to the dispersion relations. The explicit expression for the self-energy $\Pi (\omega, |\vec{k}|)$ for the $B$ meson doublet (\bplus, \bzero), arising from the interaction of eqn.(\ref{L_Bmesons}), is:
\begin{eqnarray}
\Pi (\omega, |\vec k|)  &= & \ \frac{-1}{4 \fb^2} 
\Big[3 (\rho_p +\rho_n) \pm (\rho_p -\rho_n) + 2\rho_{\gL} 
+ 2\rho_{\gS^{0}}
+  2\left( \rho_{\gS^{+}}+ \rho_{\gS^{-}}\right) \pm 
2\left(\rho_{\gS^{+}}- \rho_{\gS^{-}}\right)
\nonumber\\
&+&  \left( \rho_{\Xi^{0}}+ \rho_{\Xi^{-}}\right) \pm 
\left(\rho_{\Xi^{0}}- \rho_{\Xi^{-}}\right) \Big] \omega 
+  \frac {m_B^2}{2\fb} (\gs' +\sqrt 2 {\gz_b} ' \pm \gd') \nonumber\\
&+ & \Bigg[ \frac {d_1}{2 \fb^2} \big(\rho_p ^s +\rho_n ^s 
+ \rho_{\Lambda}^s+\rho_{\Sigma^{+}}^s 
+\rho_{\Sigma^{0}}^s + \rho_{\Sigma^{-}}^s 
+\rho_{\Xi^{0}}^s+\rho_{\Xi^{-}}^s\big) \nonumber\\
&+ & \frac{d_2}{4 \fb^2} \Big( 3 (\grssub{p} +\grssub{n}) 
\pm (\grssub{p} -\grssub{n}) + 2\grssub{\gL} 
+ 2\left( \grssub{\gS^{+}}+ \grssub{\gS^{-}}\right) \pm
2\left(\grssub{\gS^{+}}- \grssub{\gS^{-}}\right)\nonumber\\
&+ & 2\grssub{\gS^{0}} + \left( \grssub{\Xi^{0}}+ 
\grssub{\Xi^{-}}\right) \pm \left(\grssub{\Xi^{0}}- 
\grssub{\Xi^{-}}\right) \Big) 
-\frac{1}{\fb}(\gs' +\sqrt{2} \zeta_b' \pm \gd') 
\ \Bigg](\go^2 - |{\vec k}|^2)
\label{selfenergy_BplusB0}
\end{eqnarray}
where the $+$ and $-$ signs refer to the \bplus \ and \bzero \ mesons, respectively. 
Likewise, for the  $\bar{B}$ meson doublet (\bminus, \bbarzero), the expression for self-energy is given as, 
\begin{eqnarray}
\Pi (\omega, |\vec k|)  &= & \ \frac{1}{4 \fb^2} 
\Big[3 (\rho_p +\rho_n) \pm (\rho_p -\rho_n) + 2\rho_{\gL} 
+ 2\rho_{\gS^{0}}
+ 2\left( \rho_{\gS^{+}}+ \rho_{\gS^{-}}\right) \pm
2\left(\rho_{\gS^{+}}- \rho_{\gS^{-}}\right)\nonumber\\
&+ & \left( \rho_{\Xi^{0}}+ \rho_{\Xi^{-}}\right) 
\pm \left(\rho_{\Xi^{0}}- \rho_{\Xi^{-}}\right) \Big] \omega 
+ \frac {m_B^2}{2\fb} (\gs' +\sqrt 2 {\gz_b} ' \pm \gd') \nonumber\\
&+ & \Bigg[ \frac {d_1}{2 \fb^2} \big(\rho_p ^s +\rho_n ^s + 
\rho_{\Lambda}^s+\rho_{\Sigma^{+}}^s 
+\rho_{\Sigma^{0}}^s + \rho_{\Sigma^{-}}^s +\rho_{\Xi^{0}}^s
+\rho_{\Xi^{-}}^s\big) \nonumber\\
&+ & \frac{d_2}{4 \fb^2} \Big( 3 (\grssub{p} +\grssub{n}) 
\pm (\grssub{p} -\grssub{n}) + 2\grssub{\gL}
 + 2\left( \grssub{\gS^{+}}+ \grssub{\gS^{-}}\right) \pm
2\left(\grssub{\gS^{+}}- \grssub{\gS^{-}}\right)\nonumber\\
&+ & 2\grssub{\gS^{0}} + \left( \grssub{\Xi^{0}}+ 
\grssub{\Xi^{-}}\right) \pm \left(\grssub{\Xi^{0}}- 
\grssub{\Xi^{-}}\right) \Big) 
-\frac{1}{\fb}(\gs' +\sqrt{2} \zeta_b' \pm \gd') 
\ \Bigg](\go^2 - |{\vec k}|^2),
\label{selfenergy_BminusBbar0}
\end{eqnarray}
where, once again, the $+$ and $-$ signs refer to \bminus \ and \bbarzero \ mesons, respectively. 
Additionally, we also study the optical potentials of these \BandBbarm s in the
present investigation, which are defined as,
\begin{equation}
U(\go, k) = \go(k) - \sqrt{k^2 + \mb^2}
\label{OptPot_Def}
\end{equation}
where $k \ (= |{\vec k}|)$, is the momentum of the respective meson, and $\go(k)$ refers to its momentum dependent medium mass, obtained from the dispersion relation.

In the present work, we study the in-medium masses and the optical potentials
of the $B$ and $\bar B$ mesons, which arise through the scalar and number 
densities of the baryons as well as the medium dependence of the scalar fields  
at given baryon density and temperature of the hadronic matter. 
The density and temperature dependence of the scalar fields
are obtained by solving the coupled equations of motion of the scalar and vector fields. 
We also study the sensitivity of the medium modifications of these mesons,
due to the isospin asymmetry as well as strangeness of the hadronic matter,
measured by 
the isospin asymmetry parameter, $\eta$ 
and strangeness fraction, $f_s$. These parameters are defined as,
$\gn \ = -\sum_{i} ( I_{3i} \rho_{i})/\grb$,
where $I_{3i}$ denotes the (third) $z$ - component of 
isospin of the $ i^{th}$ baryon,
and, $f_s = \sum_{i}  (|S_{i}| \rho_{i})/\rho_{B}$, 
where $S_{i}$ is the strangeness 
quantum number of the $i^{th}$ baryon. A detailed account of the dependence of the 
in-medium mass of the \BandBbarm s on these parameters is given in the next section.

\end{section}

\begin{section}{Results and Discussion} \label{results_section}

Before embarking on a description of what the above formulation entails 
for the in-medium properties of the \Bm s in a hadronic environment, 
we first mention about the choice of parameters. 
The parameters of the chiral model are fitted to the vacuum masses 
of baryons, the nuclear saturation properties and other vacuum 
characteristics within the mean field approximation 
\cite{Pap_prc99,Zsch}. In this investigation, we employ 
the same parameter set that has earlier been used to study 
kaons properties in hyperonic matter \cite{mam_kaons2008}
as well as the open charmed ($D$) mesons within this chiral 
model framework \cite{arvDepja}, and refer the interested reader to 
Refs.\cite{Pap_prc99,mam_kaons2008,arvDepja} for a detailed account of these 
fitting procedures. 
The parameters $d_1$ and $d_2$ are
fitted to the empirical values of the low energy kaon-nucleon 
scattering lengths in the I=0 ans I=1 channels 
\cite{mam_kaons2008,sambuddha1,sambuddha2}.
For a further extension to the bottom sector, 
the only additional parameter required is the \Bm \ decay constant, 
\fb, for which we use the value $190.6$ MeV, consistent with the 
latest PDG \cite{PDG2012}. We observe however, that this value 
is obtained \cite{PDG2012} simply as an average over lattice QCD 
results \cite{LQCD1_fB,LQCD2_fB}. The value of $f_B$ 
obtained using QCD Sum Rules \cite{QSR1_fB, QSR2_fB, QSR3_fB} is  
slightly higher ($\sim 207$ MeV), whereas the value
of the $B$-decay constant typically taken in the literature
\cite{LQCD1_fB,LQCD2_fB,ETM_fB,fB_Lucha, LQCD3_fB}
lies in the range of around 186 MeV to 197 MeV. 
We expect that the choice of a slightly different 
value of \fb \ will 
not make much difference to the results of the present investigation.  

In Ref. \cite{roder}, the behavior of quark condensates was studied 
for various $U(N_f)_L \times U(N_f)_R$ linear sigma models, 
within the Hartree approximation, derived using the 
Cornwall-Jackiw-Tomboulis (CJT) formalism. 
Here, $N_f$ denotes the number of quark flavors. 
Their model-independent analysis of the behavior of the light 
($N_f = 2$), strange ($N_f = 3$) and charmed ($N_f = 4$) quark 
condensates is invaluable for the generalization to another additional 
flavor, as is being considered here. It is observed that while there 
are substantial changes in the light quark condensate, and a comparatively 
more subdued variation of the strange quark condensate, there is a 
near-constancy of the charmed quark condensate up to a temperature 
of around $200$ MeV (with some exotic variation above this temperature).
 Since the effective field theoretical model used in our investigation 
constitutes a hadronic description of matter, we expect our model 
to give a reasonable description of reality, only as long as we have 
a hadronic phase in QCD. In light of the fact that above a pesudo-critical 
temperature $T_c$ $\approx$ 170 MeV, QCD is predicted to go over 
to the deconfined regime \cite{LQCD1,LQCD2,LQCD3}, where we do not 
have hadrons, the constancy of charmed quark condensate, 
as obtained in \cite{roder}, is thus expected to be valid over 
the entire hadronic regime of QCD. 
(The value of $T_c$ quoted here originates from the calculations of the MILC collaboration \cite{LQCD2}; a more recent computation by the HotQCD collaboration \cite{LQCD3} predicts a smaller transition temperature ($T_c=154 \pm 9$ MeV), but that does not make  any qualitative difference to the above conclusion.) 
This analysis has been used in support of the neglect of medium effects for the 
charmed condensate $\left<\bar{c}c\right>$, in many works in the 
literature concerning the charmed mesons 
\cite{mamD2004, arindam, arvDprc, arvDepja}. 
This also bodes well with expectations from physical grounds, 
since the mass of the charm quark is above the typical QCD energy 
scale $(\sim 1 \ {\rm GeV})$, below which non-perturbative approaches 
like our effective hadronic \lag \ is considered appropriate 
\cite{leshouches}. However, since the mass of the bottom quark 
is still higher, it is reasonable to build on the analysis of 
Ref. \cite{roder}, and treat the bottom degrees of freedom 
to be frozen in the medium as well. Consequently, for the entire 
numerical analysis in the present investigation, we neglect the medium 
effects on the $B$ and $\bar B$ mesons due to the bottom condensate 
$\left<\bar{b}b\right>$. 
More specifically, we neglect the variation of the bottomed scalar field 
($\gz_b$) from its vacuum value and set ${\gz_b}' = \gz_b - {\gz_b}_0 \equiv 0$ in the current investigation.

\begin{figure}
\begin{center}
\scalebox{0.7}{\includegraphics{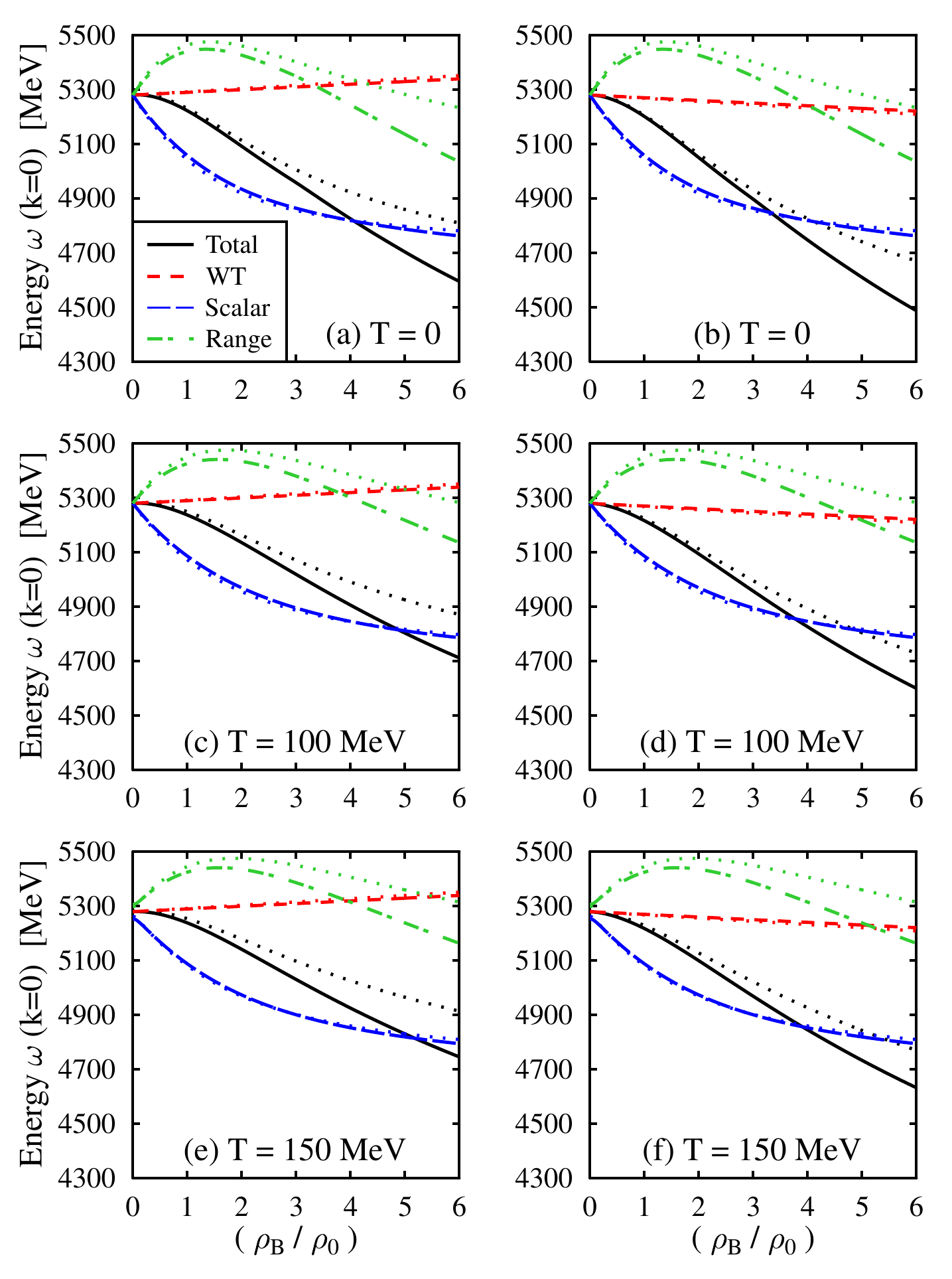}}\caption{ \label{BBbar_TbyT_eta0} (Color Online) The various contributions to the energy at ${\vec k} = 0$, for the \BandBbarm s in isospin symmetric matter ($\eta = 0$), at different temperatures. Subplots (a), (c) and (e) correspond to the 
degenerate \Bm s ($B^+$, $B^0$), while (b), (d) and (f) correspond to the 
degenerate ${\bar B}$ mesons ($B^-$, ${\bar B}^0$). 
For each case, the individual contributions in hyperonic matter (with $f_s = 0.5$), as described in the legend, are also compared against the nuclear matter situation ($f_s = 0$), represented by dotted curves.}
\end{center}
\end{figure}

We now analyze the in-medium behavior of \BandBbarm s, as per the formulation 
of the previous section. The contributions of the various individual 
interaction terms to the total in-medium masses of the \BandBbarm s 
are shown in Fig. \ref{BBbar_TbyT_eta0}, for isospin symmetric 
($\eta = 0$) nuclear ($f_s = 0$) as well as hyperonic matter, 
with $f_s = 0.5$. In symmetric matter, the scalar meson exchange term 
gives an attractive contribution to all four  of these \BandBbarm s, 
hence lowering their medium mass. This can be understood by realizing 
that since $\gz_b$ is being treated as frozen and the value of the 
scalar-isovector field $\gd$ is zero in the symmetric situation, 
the entire variation of this interaction term is due to the fluctuations 
of the scalar-isoscalar $\gs$ field. The behavior of these scalar fields 
in this chiral effective model, in both nuclear and hyperonic matter 
situations, has been studied in detail in Refs. \cite{arvDepja,arvDprc}. 
It is observed that $\gs' = \gs - \gs_0 > 0$ at all finite densities, 
hence, it follows from eqns. (\ref{selfenergy_BplusB0}) and 
(\ref{selfenergy_BminusBbar0}) that the contribution of this term 
is attractive, for the entire range of density variation considered 
here. On the other hand, the behavior of the total range term, which is
the sum of the contributions from the $d_1$ range term, $d_2$ range term 
and the first range term, is quite non-trivial. 
It is observed that the total contribution of these range terms 
is repulsive till a density of about $1.5 \grz$; thereafter, 
it becomes attractive and contributes further to a lowering 
of the in-medium mass for the \BandBbarm s. 
This kind of behavior arises 
because of the interplay of the repulsive first range range term 
and the attractive $d_1$ and $d_2$ range terms. It follows from 
their respective expressions that while the density dependence 
of the first range range term is because of the \gs \  field 
in this symmetric matter situation, that for the other two range 
terms is through scalar densities of the nucleons and the hyperons. 
While the relation $\gs' \approx \grssub{\ }$ holds approximately 
for smaller densities, at larger densities, there is considerable 
departure from this approximate equality and the density dependence 
of \gs \ is significantly sub-linear, becoming progressively more 
and more sluggish as we go to higher densities. On the other hand, 
scalar densities of all eight baryons increase monotonically with 
\grb. Thus, it is inevitable that these progressively growing 
attractive contributions, though initially smaller, would predominate 
over the decreasing magnitude of the repulsive first range term, 
which explains the observed behavior of the range terms.   

It follows from the eqn.(\ref{selfenergy_BplusB0}) that the isospin 
pair constituted by the two \Bm s (\bplus, \bzero) is degenerate 
in isospin symmetric matter, irrespective of the value of $f_s$.
(In making this assertion, we are neglecting the small $0.33$ MeV 
difference in their vacuum masses \cite{PDG2012}, since this number
is much smaller than the typical magnitude of their mass shifts that 
we are concerned with.) 
This is because symmetric matter not only has an equal number of 
isospin-pair-baryons, $(p, n), (\gS^+, \gS^-), (\gX^0, \gX^-)$, 
but also, the scalar-isovector field $\gd$ vanishes. With this, 
the asymmetric contributions to the \wtt \ and the $d_2$ range term 
vanish, while the first range term contributes equally to the isospin 
doublets, just like the scalar 
meson exchange term which was addressed before.
The $d_1$ range term is anyways common for all four mesons, 
even in the asymmetric situation, as can be seen explicitly 
from the self-energy expressions. Similar to the masses of the
$B^+$ and $B^0$, which are identical in symmetric matter,
the masses of the $B^-$ and $\bar {B^0}$ also remain same 
in isospin symmetric matter at finite densities. 
In vacuum, the masses of $B^+$ and $B^0$ coincide with the
masses of $B^-$ and $\bar {B^0}$. This, however is no longer
the case at finite densities.
For example, in cold ($T=0$) nuclear matter, the values of 
the mass drop $\Delta m = ( m_{\rm vacuum} - m_{\grb} )$ 
for the \BandBbarm s at $\grb =  \grz$ are $49$ and $74$ MeV 
respectively, which grow to $165$ and $217$ at 2$\rho_0$ and,
$357$ and $454$ MeV respectively, at $4\grz$. This difference 
arises because, the isospin symmetric 
part of the vectorial Weinberg-Tomozawa interaction has equal and opposite  
contributions for these antiparticle pairs, as can be seen explicitly 
from the expressions for their self-energies.
Being repulsive for the \Bm s (subplots (a), (c) and (e) in Fig.\ref{BBbar_TbyT_eta0}) and attractive for the \Bbarm s (subplots (b), (d) and (f)), this interaction term leads to an extra drop in the medium mass for the latter as compared to the former. 
 
As we go from symmetric nuclear to symmetric hyperonic matter, i.e. 
increase the value of $f_s$, the in-medium mass of both \BandBbarm s 
is observed to decrease. For example, the $49$ and $74$ MeV mass drops 
for the \BandBbarm s respectively, at $\grb = \grz$ in cold nuclear matter 
mentioned earlier, grow to $57$ and $78$ MeV respectively, for cold 
hyperonic matter with $f_s = 0.5$. 
The mass drops for these two sets of mesons are $198$ and $231$ at
$2\rho_0$ and, $454$ and $532$ MeV respectively at $\grb = 4\grz$ 
in the $f_s = 0.5$ situation, which 
are significantly higher than the corresponding numbers in nuclear 
matter ($165$ and $217$ for $2\rho_0$ and $357$ and $454$ MeV at 4$\rho_0$, 
as mentioned before). To understand 
this overall decrease, we must analyze the effects of increasing 
$f_s$ on each of these individual contributions to the total 
in-medium mass. It is established \cite{arvDepja} that \gs \ increases 
in magnitude with an increase in $f_s$ up to a certain density 
(e.g. $\sim 3.9\grz$ in the $T=0$ situation), beyond which 
it starts decreasing. 
As we had noted previously, the entire variation of the scalar meson exchange term is because of this \gs \ field in symmetric matter; hence, its behavior is exactly in accordance with that of \gs. In particular, it can be clearly discerned that, for the $T=0$ case for example, while the contribution from this term for the hyperonic $(f_s = 0.5)$ matter case is smaller in magnitude as compared to that for the nuclear matter for $\grb < 3.9\grz$, a reversal occurs for densities higher than this. This reversal, and especially the precise value of crossover point, is exactly in accordance with the observed behavior for the \gs \ field. The same reasoning extends over to the the magnitude of the (repulsive) first range term, though there is no such reversal for the total range term. This is so, because of the contributions from the $d_1$ and $d_2$ range terms, which are totally dependent on the scalar densities. As was reasoned already, the decreasing magnitude of the \gs \ dependent repulsive contribution makes it inevitable that the large density behavior of the total range terms would be governed by these attractive, \grssub{\ }-dependent interaction terms. Now, the effect of increasing $f_s$ on the $d_1$ range term is to increase the magnitude of the attractive interactions at larger densities. This result appears surprising at first glance, since this $d_1$ range term depends only on the sum of all eight scalar densities, $(\sum_i \grssub{i})$. By increasing $f_s$, we are only redistributing the total density amongst these eight species, so one may naively expect the sum to remain the same irrespective of the value of $f_s$. However, this is not the case, since this process redistributes the total number density $(\grb = \sum_i \grsub{i})$, and not the scalar density, for which there is no such overall conservation. While $\grssub{B} \approx \grsub{B}$ at small densities, at larger densities, the former, though still an increasing function, has much more subdued increase with respect to the latter. The fact that they are identical at small densities, implies an approximate overall conservation at small densities, in line with the above reasoning, which is indeed observed to be the case. Thus, while this interaction term has a negligible $f_s$ dependence till $\grb \approx \grz$, there is a considerable increase in its magnitude at larger densities, with an increase in $f_s$. 
On the other hand, the $f_s$ dependence of $d_2$ range term is comparatively less pronounced, and has the exactly opposite behavior - this term is observed to increase with $f_s$ till about $5\grz$ and show a marginal increase thereafter. 
Naturally then, the larger magnitude of the $d_1$ range term dominates over the other two, and the overall behavior of range terms is to decrease with $f_s$. This 
is especially pronounced at larger densities, where it is responsible for 
causing a still larger decrease in the medium mass for the \BandBbarm s. 
Lastly, the \wtt \ is known to decrease in magnitude as the value of 
$f_s$ is increased, which can be inferred by repeating the argument, 
presented for $D(\bar D)$ mesons in Ref. \cite{arvDepja}. 
It follows from eqns. (\ref{selfenergy_BplusB0}) and (\ref{selfenergy_BminusBbar0}), that in symmetric nuclear matter, the magnitude of this interaction term is proportional to $(3(\grsub{p}+\grsub{n}))$, which would imply a precisely linear increase (or decrease, for the corresponding antiparticle) with density, since $\grb = (\grsub{p}+\grsub{n})$ in nuclear matter. However, in symmetric hyperonic matter, the total \grb \ is redistributed amongst all eight baryons, such that the magnitude of this interaction term is proportional to $(3(\rho_p +\rho_n) + 2\left( \rho_{\gL} + \rho_{\gS^{0}} + \rho_{\gS^{+}}+ \rho_{\gS^{-}}\right) + \rho_{\Xi^{0}}+ \rho_{\Xi^{-}})$,
which can be rearranged in a more illuminating form, as 
$\left( 3\sum_i \grsub{i} - ( \rho_{\gL} + \rho_{\gS^{0}} + \rho_{\gS^{+}}+ \rho_{\gS^{-}} + 2(\rho_{\Xi^{0}}+ \rho_{\Xi^{-}}))\right)$. Recognizing that the first term  itself equals \grb, this factor is of the form $(3\grb - g(\rho, f_s))$, where the term $g$, being totally dependent on the hyperonic number densities, is an increasing function of both $f_s$ and \grb. It follows then, that the magnitude of this \wtt \ decreases with $f_s$ for fixed \grb, and with \grb \ for fixed $f_s$, which is in exact agreement with what one observes from Fig.
\ref{BBbar_TbyT_eta0}.

Naturally then, the cumulative effect of all that has been reasoned above, is to decrease the in-medium mass of these four mesons, in hyperonic matter as compared to nuclear matter. 
Additionally, it may be observed from Fig. \ref{BBbar_TbyT_eta0} that while \BandBbarm s are still non-degenerate, the magnitude of mass-asymmetry between antiparticles reduces with $f_s$. 
For example, 
the values of $(m_B, \ m_{\bar B})$ 
in cold hyperonic matter at $\grb = \grz$, are $(5222, \ 5201)$,
$(5091,5049)$ and $(4825,\ 4747)$ MeV respectively, 5
at $\grb = \grz$, $\rho_B=2\rho_0$ and $4\grz$ respectively 
(hence implying $B--\bar B$ mass difference of $21$, $42$ and $78$ MeV, 
respectively), as against the values of $(5230,\ 5205)$, $(5112,5062)$,  
and $(4922,\ 4825)$ MeV, respectively, ($B--\bar B$ mass difference 
$25$, $50$ and $97$ MeV, respectively) in cold nuclear matter. 
This follows immediately from the reasoning of the previous paragraph, since we had noted earlier that it is the \wtt \ that is responsible for a mass shift asymmetry between particle
and antiparticle. Thus, a decrease in the magnitude of this interaction term with $f_s$, ought to have this effect.

To conclude our discussion of symmetric matter, we depart from our 
treatment of cold ($T=0$) matter and consider the effect of a finite 
temperature on the in-medium mass of these mesons. It is observed 
that respective magnitudes of the mass drops for both \BandBbarm s 
decrease at larger temperatures. For example, in symmetric nuclear 
matter, the $T=0$ mass drops of $167$ and $217$ at $\rho_B=2\rho_0$,
and, $357$ and $454$ MeV at 
$\grb = 4 \grz$ mentioned earlier for the \BandBbarm s respectively, 
shrink to $117$ and $168$ for $2\rho_0$ and  $289$ and $387$ MeV 
for $4\rho_0$, respectively at $T=100$ MeV, and 
further to $101$ and $152$ for $\rho_B=2\rho_0$, and,
 $253$ and $353$ MeV for $\rho_B=4\rho_0$ at $T=150$ MeV. Likewise, 
for strange hadronic matter with $f_s = 0.5$, the corresponding 
$T=0$ mass drops of $188$ and $250$ at $\rho_B=2\rho_0$, and,
$454$ and $532$ MeV at $\grb = 4 \grz$, are observed
to reduce to $144$ and $186$ for $\rho_B=2\rho_0$, and,
$374$ and $453$ MeV respectively at $\rho_B=4\rho_0$,
at $T=100$ MeV, and 
further to $138$ and $180$ for $\rho_B=2\rho_0$ and,
$354$ and $434$ MeV for $\rho_B=4\rho_0$, at $T=150$ MeV. This 
decrease in the mass shifts as the temperature is raised, 
originates due to a decrease in the magnitudes of the
scalar fields with increase in temperature. The weakening of the
medium effects with an increase in temperature has earlier been
observed for kaons and antikaons \cite{mamK2004} and $D$ mesons 
\cite{mamD2004, arvDprc, arvDepja} within the chiral effective model.

Everything considered so far in this discussion pertained to 
isospin-symmetric matter $(\eta = 0)$. We now proceed to discuss 
the behavior of \BandBbarm s in the more general 
situation of isospin asymmetric matter. Figures \ref{BplusB0_TbyT_eta5} 
and  \ref{BminusBbar0_TbyT_eta5} show the behavior of in-medium mass 
of the \Bm s (\bplus,\ \bzero) and the \Bbarm s (\bminus,\ \bbarzero) 
respectively, in both nuclear and hyperonic matter ($f_s = 0.5$) 
situations, along with the various individual contributions to the 
total in-medium mass of the \BandBbarm s, in asymmetric matter 
with $\eta = 0.5$.  The particles constituting the isospin doublets 
are 
observed to be 
non-degenerate in isospin-asymmetric matter. This disparity 
is evident in figures \ref{BplusB0_TbyT_eta5} and  
\ref{BminusBbar0_TbyT_eta5}, from which one can easily see that 
the mass of \bzero \ meson drops more than that of \bplus \ meson, 
and that of \bbarzero \ meson drops more than the \bminus \ meson. 
This behavior originates from the fact that the asymmetric contributions 
to the in-medium interactions, which are now non-zero, distinguish 
between these isospin pairs. 
For example, in the nuclear matter situation, the \wtt \ has an 
extra asymmetric contribution of $\pm (\grsub{p}-\grsub{n})$, 
which makes this interaction term more repulsive for \bzero \ meson 
as compared to the \bplus \ meson. Likewise, this additional term 
makes this term more attractive for \bbarzero \ meson as compared 
to \bminus \ meson, hence decreasing further the in-medium mass 
of the \bbarzero \ meson. In hyperonic matter, this \wtt \ also 
has asymmetric contributions from similar terms dependent on the 
hyperonic number densities. Due to exactly the same structure of 
interaction terms, albeit in terms of scalar densities, a similar 
reasoning applies to the $d_2$ range term. Additionally, the respective
 contributions of the scalar meson exchange term and the (repulsive) 
first range term also differ for these isospin pairs, owing to a 
$\left(\gs \pm \gd \right)$ structure in the interaction terms, 
as can be seen from eqns. (\ref{selfenergy_BplusB0}) and 
(\ref{selfenergy_BminusBbar0}). In fact, even the $d_1$ range term, 
which appears to be completely isospin symmetric (since it is 
proportional to $\sum_i \grssub{i}$), gets altered in magnitude 
as compared to the $\eta = 0$ situation. 
This difference arises because the values of the scalar fields 
calculated with $\gd = 0$ in the symmetric situation, turn out 
to be different from those calculated in the asymmetric situation, 
with $\gd \ne 0$. These values of the scalar fields are then used 
to calculate the scalar densities, which are, thus, altered between 
these two cases. Thus, it can be inferred that the entire behavior 
of the individual terms, and hence, also their interplay, is entirely 
different in isospin-asymmetric situation, as compared to the symmetric case.

\begin{figure}
\begin{center}
\scalebox{0.6}{\includegraphics{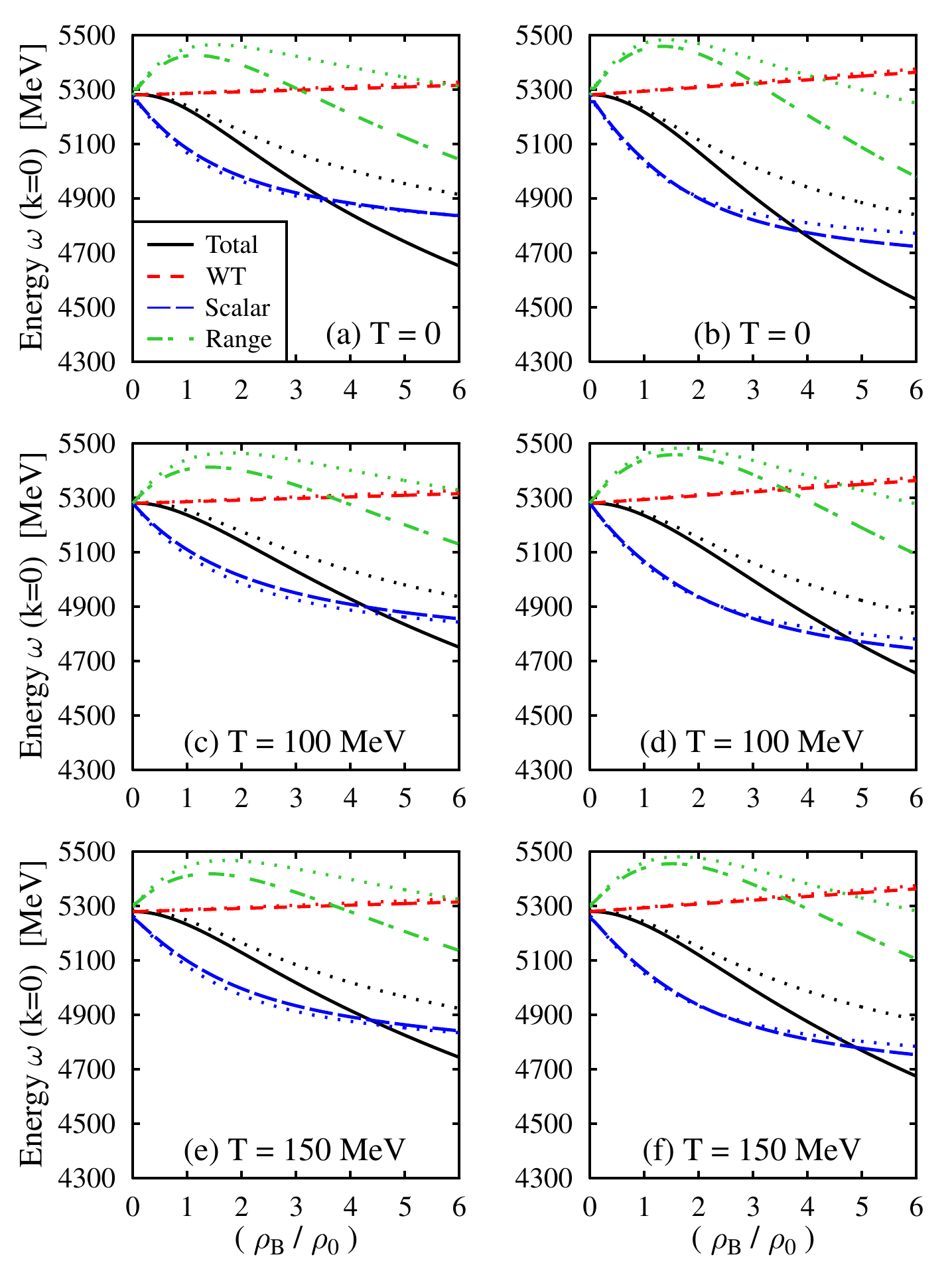}}\caption{\label{BplusB0_TbyT_eta5} (Color Online) The various contributions to the energy at ${\vec k} = 0$, for the \Bm \ doublet (\bplus, \bzero) in isospin asymmetric matter ($\eta = 0.5$), at different temperatures. Subplots (a), (c) and (e) correspond to the \bplus \ meson while (b), (d) and (f) correspond to the \bzero \ meson. For each case, the individual contributions in hyperonic matter (with $f_s = 0.5$), as described in the legend, are also compared against the nuclear matter situation ($f_s = 0$), represented by dotted curves.}
\end{center}
\end{figure} 

\begin{figure}
\begin{center}
\scalebox{0.6}{\includegraphics{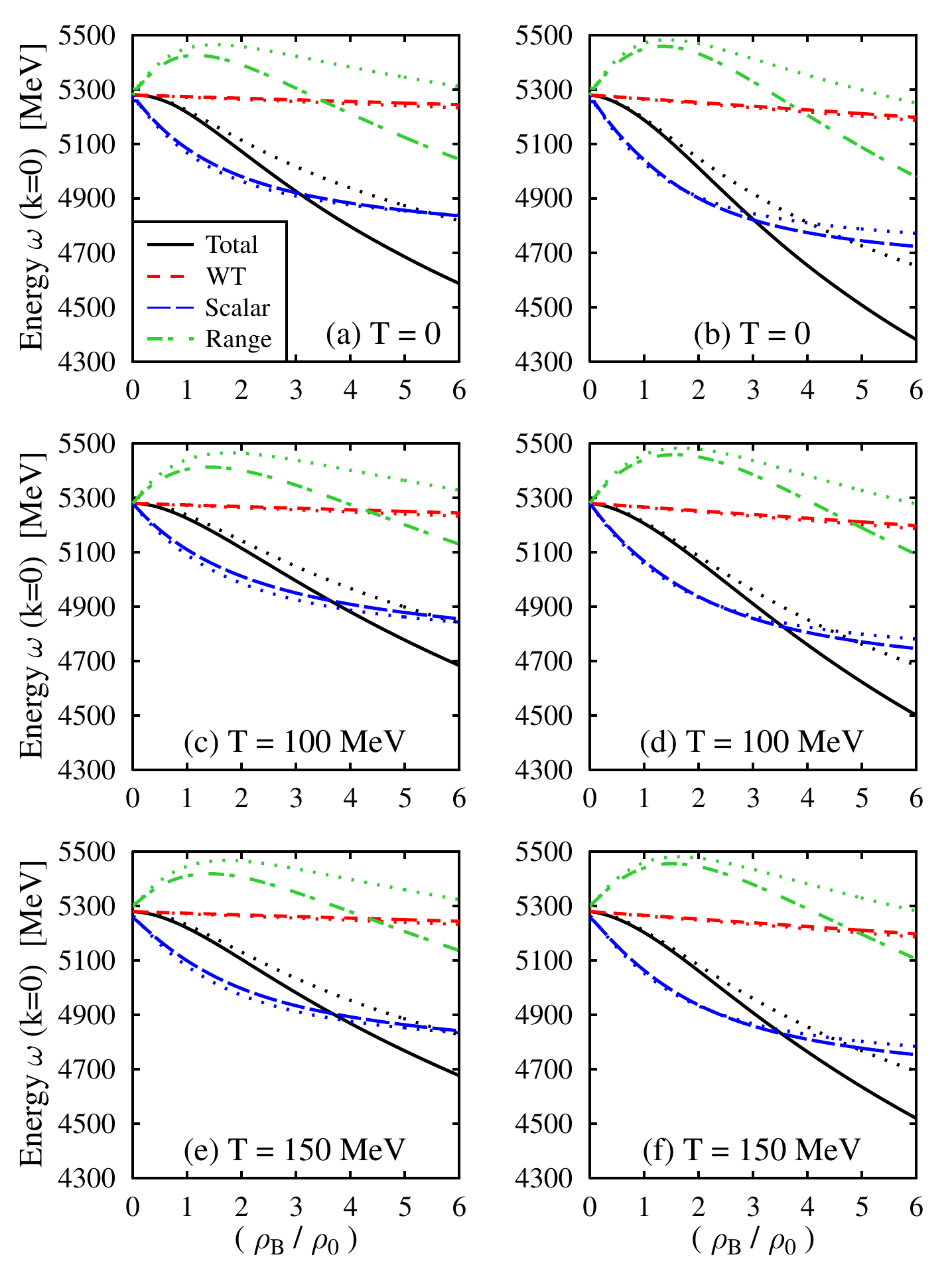}}\caption{ \label{BminusBbar0_TbyT_eta5} (Color Online) The various contributions to the energy at ${\vec k} = 0$, for the \Bbarm \ doublet (\bminus, \bbarzero) in isospin asymmetric matter ($\eta = 0.5$), at different temperatures. Subplots (a), (c) and (e) correspond to the \bminus \ meson while (b), (d) and (f) correspond to the \bbarzero \ meson. For each case, the individual contributions in hyperonic matter (with $f_s = 0.5$), as described in the legend, are also compared against the nuclear matter situation ($f_s = 0$), represented by dotted curves.}
\end{center}
\end{figure}

Thus, realizing that even in this most general situation, the behavior of the in-medium
mass of \BandBbarm s can be understood by studying this interplay of individual contri-
butions, we proceed to discuss the cumulative sensitivity of their medium mass on each of
these four parameters ($\grb, T, \eta, f_s$), as is shown in figures \ref{BplusB0_fs0fs3_etadep} - \ref{BminusBbar0_fs0fs5_etadep}. These show
the comparative behavior of these mesons in both nuclear and hyperonic matter, in both
symmetric and asymmetric situations, as a function of density, and at various temperatures. 
In order to make the effects of asymmetry and strangeness clearer from the outset, we have concerned ourselves with rather extreme values of these parameters in figures \ref{BBbar_TbyT_eta0} - \ref{BminusBbar0_TbyT_eta5}, where the only values of these parameters considered were $0$ and $0.5$. However, the physical situation in typical experimental situations is comparatively, 
more modest. For example, the isospin asymmetry in collision experiments involving lead (${\ }^{207} {\rm Pb}_{82}$) and gold (${\ }^{197} {\rm Au}_{79}$) nuclei may be estimated from the isospin asymmetry in these nuclei themselves, to a first approximation. The same comes out to be roughly $\eta \approx 0.2$ for both cases. Likewise, since the hyperons are more massive as compared to nucleons, 
it is not unrealistic at all to expect them to be less abundant than the latter in typical situations, which implies that the possibility of a 
less extreme value of $f_s$ should also be entertained. Since each of these parameter values are fed in as inputs into our calculations of the medium mass, these intermediate cases can be similarly covered, and are also included in the discussion that follows. 
While everything reasoned thus far in this article, be it the fact 
that the mass drops intensify on the addition of hyperons in the medium, 
or that (in general) the mass drops weaken at higher temperatures, 
or the degeneracy and degeneracy breaking inferred already, 
is nicely reflected in 
figures \ref{BplusB0_fs0fs3_etadep} - \ref{BminusBbar0_fs0fs5_etadep}, 
these make explicit the effect of isospin asymmetry on the in-medium 
mass of the \BandBbarm s. Since we had clamped $\eta$ to a fixed value
even while addressing their effective mass in the asymmetric situation 
in figures \ref{BplusB0_TbyT_eta5} and  \ref{BminusBbar0_TbyT_eta5}, 
we now analyze the effect of increasing $\eta$ on the in-medium mass 
of these mesons. 
We begin with the hyperonic matter ($f_s = 0.3$ situation in figures 
\ref{BplusB0_fs0fs3_etadep} - \ref{BminusBbar0_fs0fs3_etadep}, and 
$f_s = 0.5$ situation in figures \ref{BplusB0_fs0fs5_etadep} 
- \ref{BminusBbar0_fs0fs5_etadep}) and address 
the somewhat anomalous nuclear matter situation a little later. 
One can readily observe that in hyperonic matter, for each pair 
of isospin doublets, the effect of increase in asymmetry is opposite
 - producing a decrease in mass for one meson, and an increase 
for the other. This is absolutely consistent with what we have already
 observed in the individual terms - the  opposite nature of asymmetric
 contributions for \bplus \ and \bzero\ meson (as well as for the 
corresponding \Bbarm s (\bminus \ and \bbarzero)). While the symmetric
 parts of each of these individual contributions is common for these 
isospin pairs and contribute to identical decrease with density, 
these asymmetric contributions in \wtt, $d_2$ range term, as well as 
the $(\gs' \pm \gd')$ structure in the first range term as well as 
the scalar meson exchange term, are responsible for causing a further 
drop for the \bzero \ and \bbarzero \ mesons, and a relative increase
 in the medium mass for \bplus \ and \bminus, as compared to the 
symmetric situation. Further, it is observed that, in general, 
this isospin dependence decreases at larger temperatures, which 
is due to a decrease in the magnitude of $\delta$ at larger 
temperatures, hence producing a decrease in the magnitude 
of each of these asymmetric contributions. 
In nuclear matter, however, it is observed that while isospin asymmetry produces an increase and decrease, respectively, in the \bminus \ and \bbarzero \ meson mass (just like the hyperonic matter situation), the \Bm \ doublet has an apparently anomalous behavior. Here, the \bplus \ meson mass is observed to increase with asymmetry, the magnitude of isospin dependence decreasing with temperature, just like the finite $f_s$ situation. However, the \bzero \ meson mass is observed to show a small increase with asymmetry in the $T=0$ situation, while at larger temperatures, one encounters a reversal in this behavior, with the effective mass exhibiting a small decrease with asymmetry. It may be noted in particular that the magnitude of temperature dependence of \bzero \ mass is the weakest out of all four mesons, due to this gradual reversal in sign at intermediate temperatures. This apparent discrepancy is once again due to a delicate interplay between the consistently attractive scalar meson exchange term contribution, the consistently repulsive \wtt \ contribution and the contribution from the range terms which switch from repulsive to attractive at larger densities. At smaller temperatures, in the asymmetric situation, one observes that these contributions almost counterbalance each other at small densities, hence producing a resultant of small magnitude. 
At larger densities, aided by the extra asymmetric terms, the contribution from the \wtt \ is observed to dominate over the other (attractive) contributions, hence producing a net increase in mass with asymmetry.
At larger temperatures, however, the magnitude of scalar field $\gs$ increases,
while the magnitude of $\gd$ decreases, which implies that both the fluctuations
in $\gs$ and $\gd$ decrease in magnitude, hence disturbing this delicate 
balance. This has a larger effect on the attractive scalar meson exchange term and total range term, as compared to the relatively robust \wtt, with the result 
that 
the overall magnitude of the attractive terms increases over that of the repulsive contributions, producing a net decrease in mass with asymmetry at higher temperatures.     
Additionally, it can be discerned that between these four mesons, \bbarzero \ meson suffers the largest magnitude of mass drop in asymmetric hyperonic matter situation. This follows from the fact that, as has already been mentioned that
\Bbarm s have a larger mass drop as compared to \Bm s, which intensifies further in hyperonic matter situation. Adding to this the fact that \bzero \ and \bbarzero \ masses drop further in asymmetric matter, it follows that for \bbarzero \ meson, the effect of decrease in medium mass with $f_s$ gets accentuated by a decrease with isospin asymmetry parameter, hence producing the acute mass drop 
for \bbarzero, as compared to the other mesons ($B^\pm$, $B^0$). 

\begin{figure}
\begin{center}
\scalebox{0.6}{\includegraphics{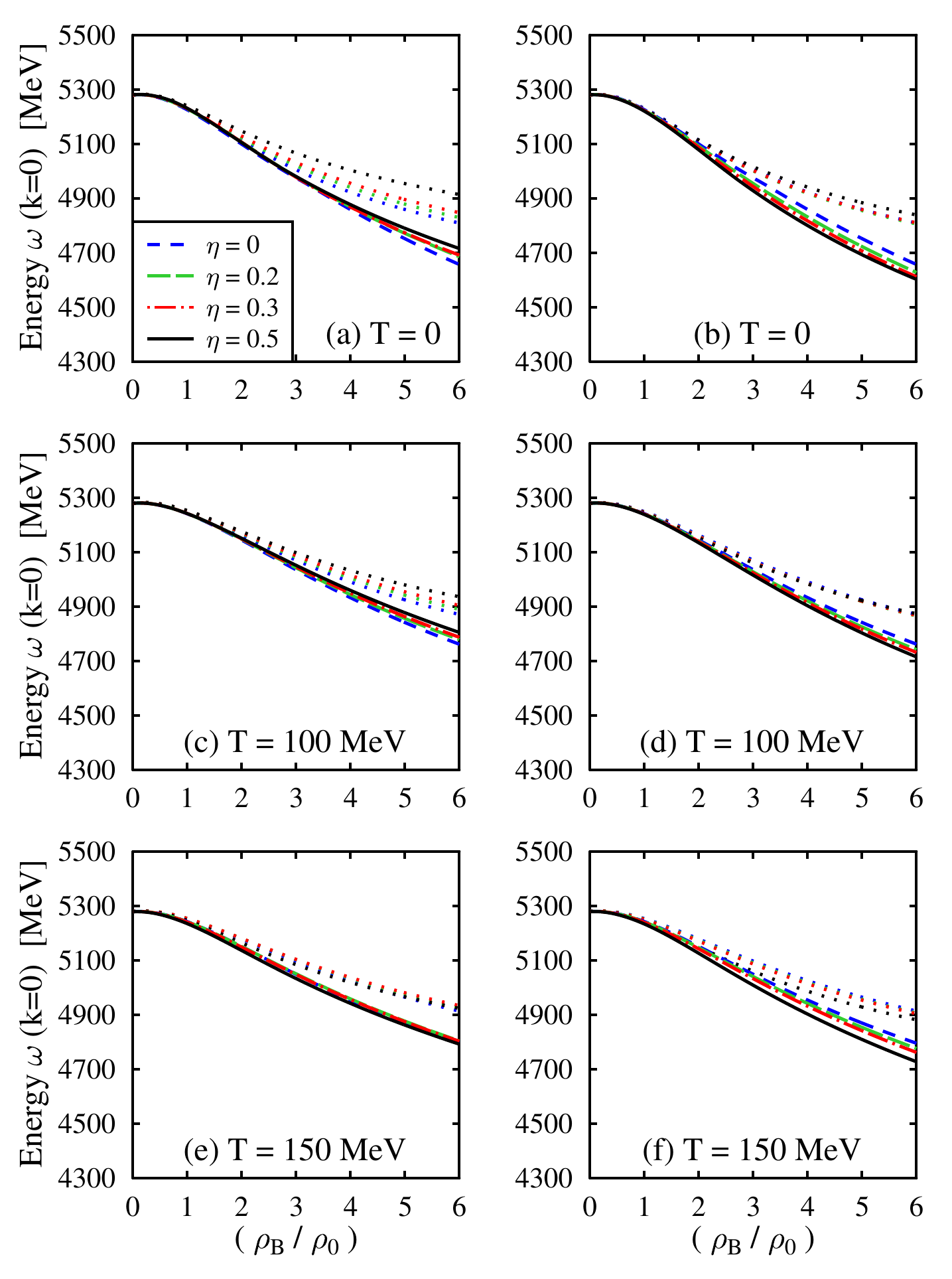}}\caption{ \label{BplusB0_fs0fs3_etadep} (Color Online) A comparison of the energy at ${\vec k} = 0$, of the \bplus \ (subplots (a), (c) and (e)) and \bzero \ mesons (subplots (b), (d) and (f)), 
%
in hyperonic matter (with $f_s = 0.3$), at various values of the isospin asymmetry parameter ($\eta$), as described in the legend, and at different temperatures. In each case, this hyperonic matter situation is also compared against the nuclear matter ($f_s = 0$) situation, represented by dotted lines.}  
\end{center}
\end{figure}

\begin{figure}
\begin{center}
\scalebox{0.6}{\includegraphics{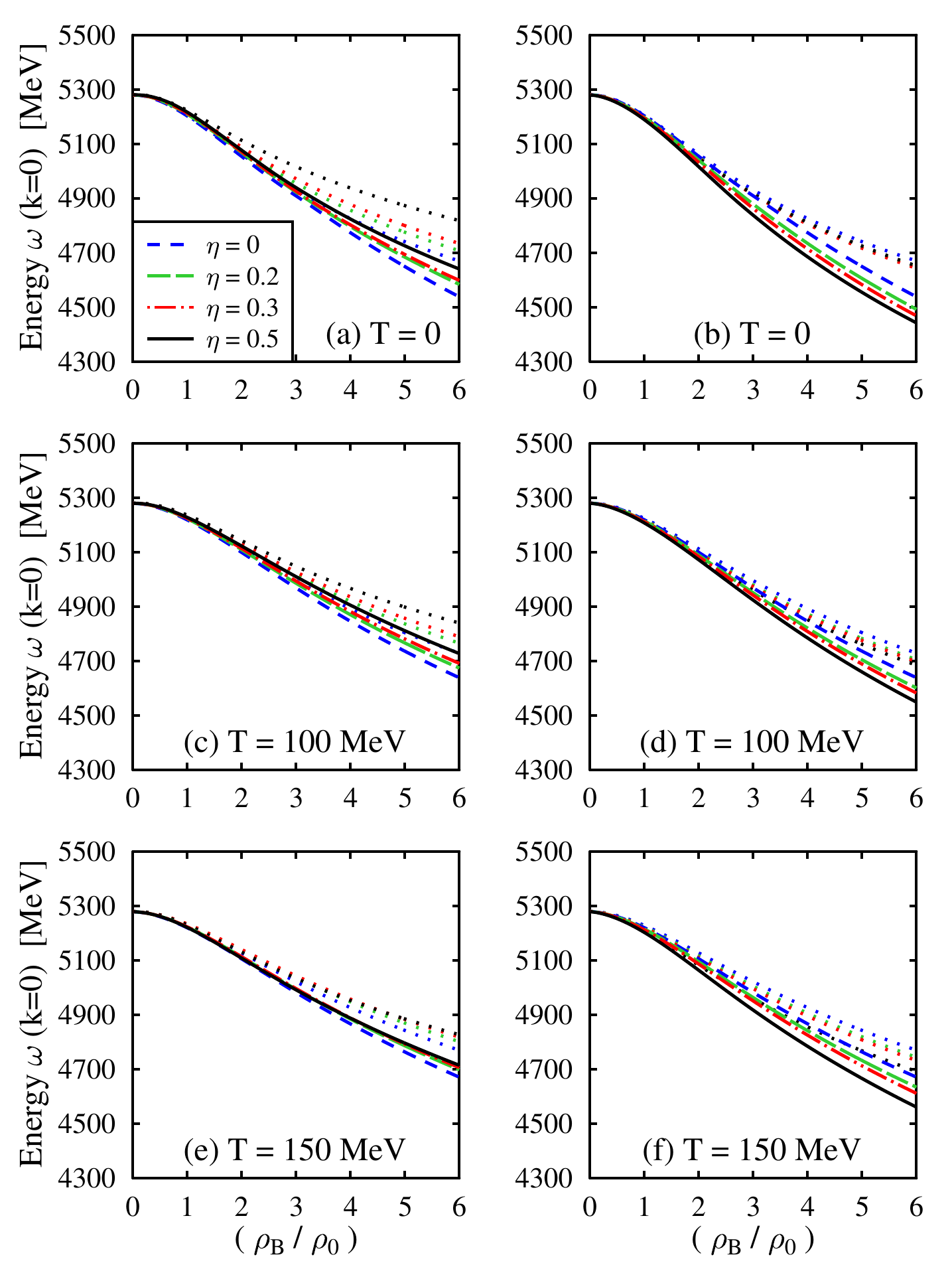}} \caption{\label{BminusBbar0_fs0fs3_etadep} (Color Online) A comparison of the energy at ${\vec k} = 0$, of the \bminus \ (subplots (a), (c) and (e)) and \bbarzero \ mesons (subplots (b), (d) and (f)), in hyperonic matter (with $f_s = 0.3$), at various values of the isospin asymmetry parameter ($\eta$), as described in the legend, and at different temperatures. In each case, this hyperonic matter situation is also compared against the nuclear matter ($f_s = 0$) situation, represented by dotted lines.}  
\end{center}
\end{figure}

\begin{figure}
\begin{center}
\scalebox{0.6}{\includegraphics{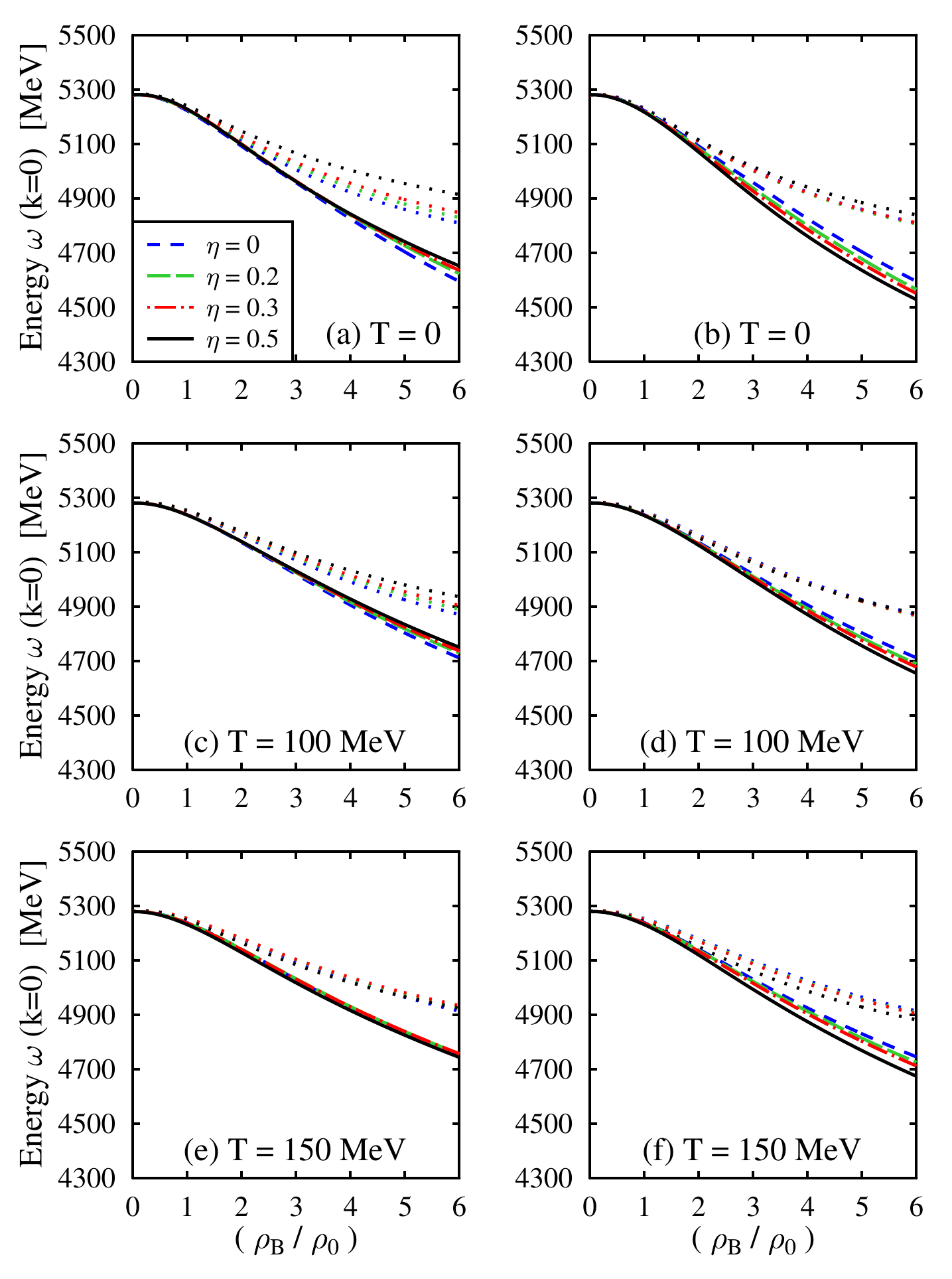}}\caption{ \label{BplusB0_fs0fs5_etadep} (Color Online) A comparison of the energy at ${\vec k} = 0$, of the \bplus \ (subplots (a), (c) and (e)) and \bzero \ mesons (subplots (b), (d) and (f)), in hyperonic matter (with $f_s = 0.5$), at various values of the isospin asymmetry parameter ($\eta$), as described in the legend, and at different temperatures. In each case, this hyperonic matter situation is also compared against the nuclear matter ($f_s = 0$) situation, represented by dotted lines.}  
\end{center}
\end{figure}

\begin{figure}
\begin{center}
\scalebox{0.6}{\includegraphics{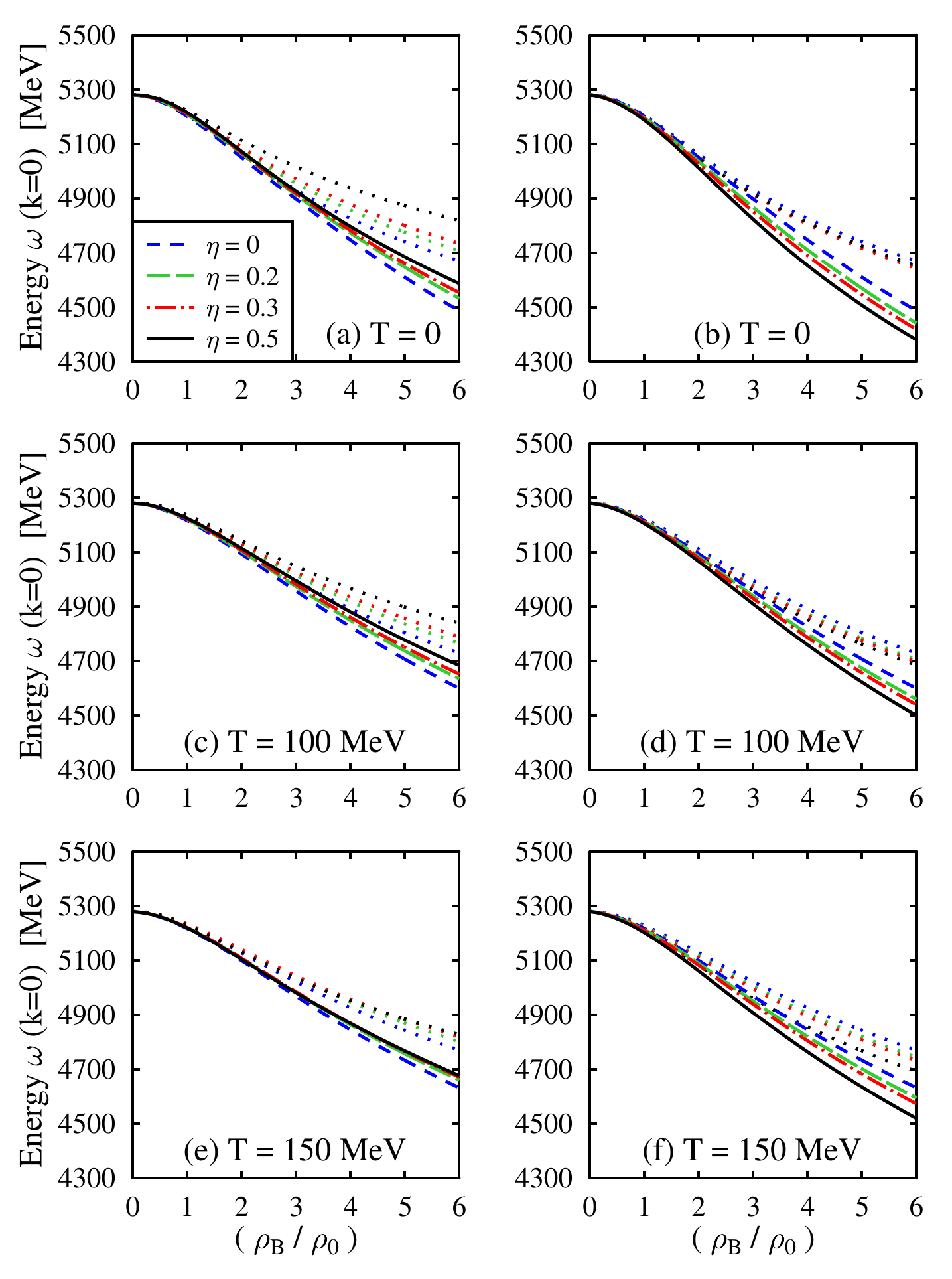}} \caption{\label{BminusBbar0_fs0fs5_etadep} (Color Online) A comparison of the energy at ${\vec k} = 0$, of the \bminus \ (subplots (a), (c) and (e)) and \bbarzero \ mesons (subplots (b), (d) and (f)), in hyperonic matter (with $f_s = 0.5$), at various values of the isospin asymmetry parameter ($\eta$), as described in the legend, and at different temperatures. In each case, this hyperonic matter situation is also compared against the nuclear matter ($f_s = 0$) situation, represented by dotted lines.}  
\end{center}
\end{figure}

As has been reasoned already, in asymmetric matter, the effect of the extra asymmetric terms is to make the \wtt \ more repulsive for \bzero \ meson as compared to the \bplus \ meson, and more attractive for the \bbarzero \ meson as compared to the \bminus \ meson. It may be noticed from the expressions of the in-medium self energies that while these extra asymmetric contributions break the mass degeneracy of isospin doublets as was seen above, the mass degeneracy of antiparticles (\bplus, \bminus) and (\bzero, \bbarzero) is still getting broken because of equal and opposite contributions of this \wtt \ only. Putting these two factors together, it follows that the magnitude of mass shift asymmetry between (\bzero, \bbarzero) is larger than that between (\bplus, \bminus), as can be seen from figures \ref{BplusB0_TbyT_eta5} and \ref{BminusBbar0_TbyT_eta5}. For example, in cold hyperonic matter, with $f_s = 0.5$, the values of $(\Delta m_{\bplus}, \Delta m_{\bminus})$ are observed to be $(52,64)$, $(182,207)$ and 
$(437,483)$ MeV, at $\grb = \grz$, $2\rho_0$ and $4\grz$ respectively. 
These may be compared to the numbers $(63,92)$, $(210,268)$
and $(519,626)$ MeV 
for the $(\Delta m_{\bzero}, \Delta m_{\bbarzero})$ mesons under 
the same conditions. 
We might note here that the corresponding values in the symmetric situation, 
$(57,78)$, $(188,230)$ and $(454,532)$ MeV for the B and $\bar B$ mesons, 
($\Delta m_B$, $\Delta m_{\bar B}$) at densities of  $\rho_0$, $2\rho_0$ 
and 4$\rho_0$ respectively, were identical for the two pairs, since 
both the \Bm s and both the \Bbarm s were degenerate in that situation. 
Since all other contributions, even with the extra asymmetric contributions, are exactly identical for antiparticles, this behavior is completely borne out of the larger magnitude of the contribution from this \wtt \ for the antiparticle pair (\bzero, \bbarzero) as compared to (\bplus, \bminus). Also, it clearly follows from figures \ref{BplusB0_fs0fs3_etadep} - \ref{BminusBbar0_fs0fs5_etadep} that the 
variation of medium mass for either meson with 
both \isoap \ and strangeness is completely monotonic, 
since the mass drops corresponding to intermediate values of these parameters (which, as we saw earlier, are more realistic choices from the point of view of experimental relevance) are also intermediate between the two extreme situations considered by us, for either of these parameters.

Building on this analysis of the effective mass of \BandBbarm s, 
we now depart from the line of approach followed so far in this article, 
where we have considered the medium effects at $\vec{k} = 0$, and venture 
into the finite momentum regime. To this end, we consider the in-medium 
optical potentials for the \BandBbarm s, defined via eqn.(\ref{OptPot_Def}). 
Figures \ref{OptPot_B_fs0fs3}-\ref{OptPot_Bbar_fs0fs5} show the variation of optical
potentials of 
\BandBbarm s with momentum 
$k \ (= |{\vec k}|)$, at $\grb = \grz$, $2\grz$ and $4\grz$, 
in both symmetric and asymmetric, nuclear and hyperonic matter at
$T=0$. In order to appreciate this behavior of optical potentials, we note that as per its definition (\ref{OptPot_Def}), at $k=0$, optical potential is just the negative of the mass drop of the respective meson, i.e. 
\begin{equation}
U(k=0) = - \Delta m (k=0) \equiv - \Delta m (\grb, T, \eta, f_s) ,
\end{equation} 
which has been treated in detail in this article. Thus, the behavior of the intercepts follows immediately from this realization 
-- be it the largest magnitude for \bbarzero \ between all four of these mesons, or their equivalence in the symmetric situation for the pairs (\bplus, \ \bzero) and (\bminus, \bbarzero), or the lower optical potentials for the \Bbarm s as compared to the \Bm s in the symmetric situation, etc. It may additionally be noticed from the self energy expressions, given by 
eqns. (\ref{selfenergy_BplusB0}) and (\ref{selfenergy_BminusBbar0}), that in the symmetric situation, just like everything noted earlier, the $k$ dependence for both the \Bm s (as well as both the \Bbarm s) is also identical. 
This readily explains why the curves corresponding to (\bplus, \bzero) and (\bminus, \bbarzero) mesons are identical for $\eta = 0$, even at finite $k$. 
In fact, the effects of non-zero strangeness, as well as non-zero asymmetry, as analyzed earlier in the ${\vec k} = 0$ situation, extend directly to this finite momentum situation. The former is responsible for lowering the optical potentials of all four of these mesons as compared to the nuclear matter $(f_s = 0)$ situation, while the effect of the latter is observed to be opposite for isospin pairs, except for the somewhat anomalous behavior for the \bzero \ meson, noted before. 
Also, a monotonic variation with the \isoap, as was observed with the medium masses, is reflected in the optical potentials as well.  
However, a general observation from the plots is that, the effect of 
non-zero $k$ is to lower the optical potential from its value at zero 
$k$, which follows from the definition, eqn.(\ref{OptPot_Def}). 
Both the terms, $\go(k)$ (calculated from the dispersion relation, 
with non-zero $k$) and the free kinetic energy part 
$(k^2 + m_B^2)^{1/2}$ are increasing functions of $k$. However, 
the former increases faster, since its $k$ dependence arises through 
the factor $k^2(1+ d_1 f_1(\grssub{i}) + d_2 f_2(\grssub{i}) 
- f_3(\gs',\gd'))$ against the plain $k^2$ of the latter. 
We have already discussed the behavior and interplay of these functions while studying the behavior of the total range term with density. Discounting the small density regime where the negative contribution predominates over the positive ones, the factor in parenthesis is significantly larger than $1$ at larger densities, which is responsible for the observed reduction in the magnitude of optical potential with $k$. Also, a larger reduction with $k$ at, e.g. $\grb = 4\grz$, as compared to $\grb = \grz$, is exactly in tow with the density dependence of this factor. Thus, with optical potentials falling monotonically with $k$, the largest magnitude for each of them is seen at $k=0$, which is the largest for \bbarzero \ amongst all four of them, for reasons already 
described in detail.

\begin{figure}
\begin{center}
\scalebox{0.59}{\includegraphics{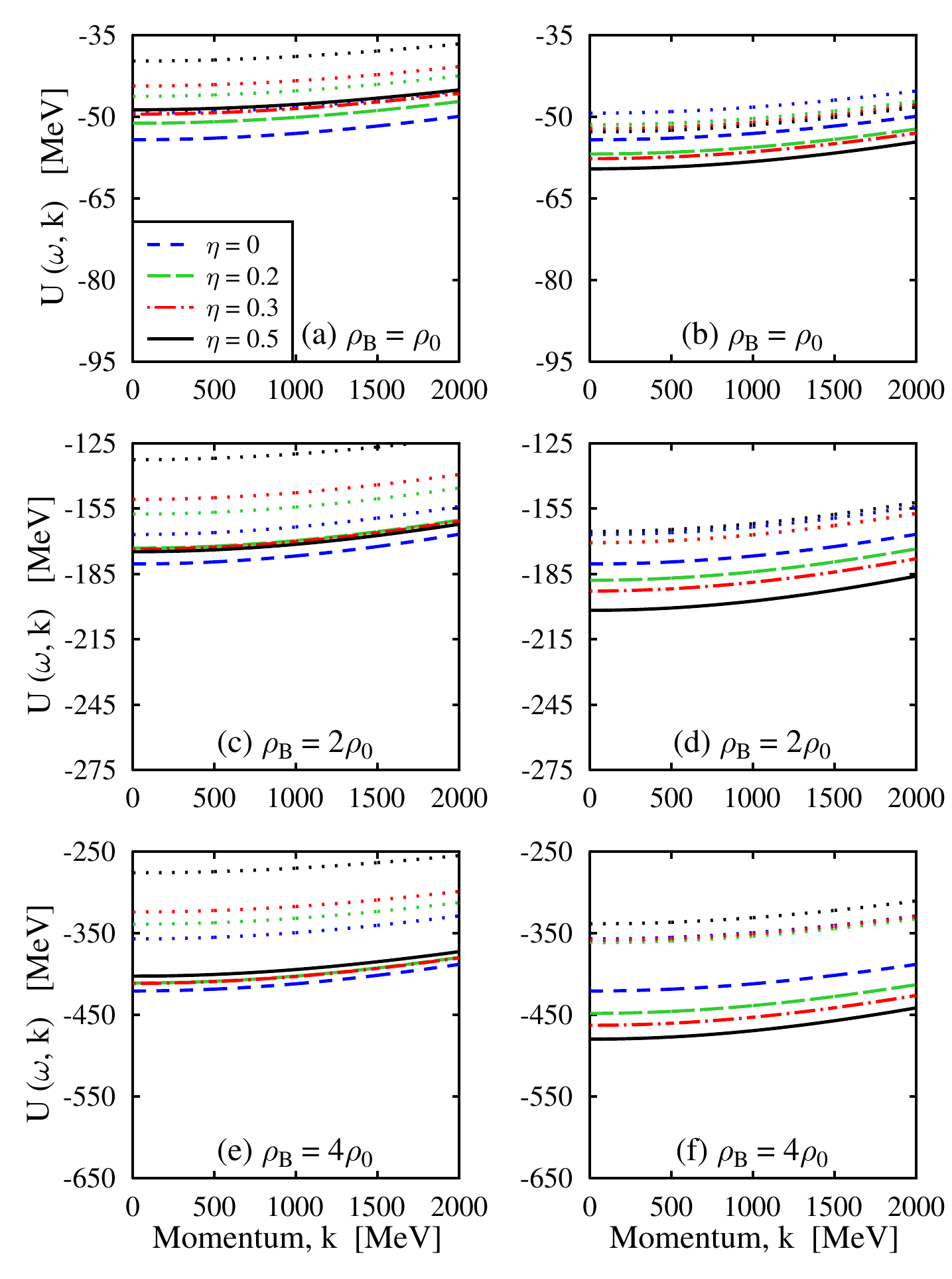}}\caption{ \label{OptPot_B_fs0fs3} (Color Online) The optical potentials of the \Bm s (\bplus \ in subplots (a), (c) and (e), and \bzero \ in subplots (b), (d) and (f)), as a function of momentum $k \ (\equiv |{\vec k}|)$, in cold $(T=0)$ hyperonic matter (with $f_s = 0.3$), at various values of the \isoap \ ($\eta$), as described in the legend, and at different 
densities. In each case, this hyperonic matter situation is also compared against the nuclear matter ($f_s = 0$) situation, represented by dotted lines.}  
\end{center}
\end{figure}
\begin{figure}
\begin{center}
\scalebox{0.59}{\includegraphics{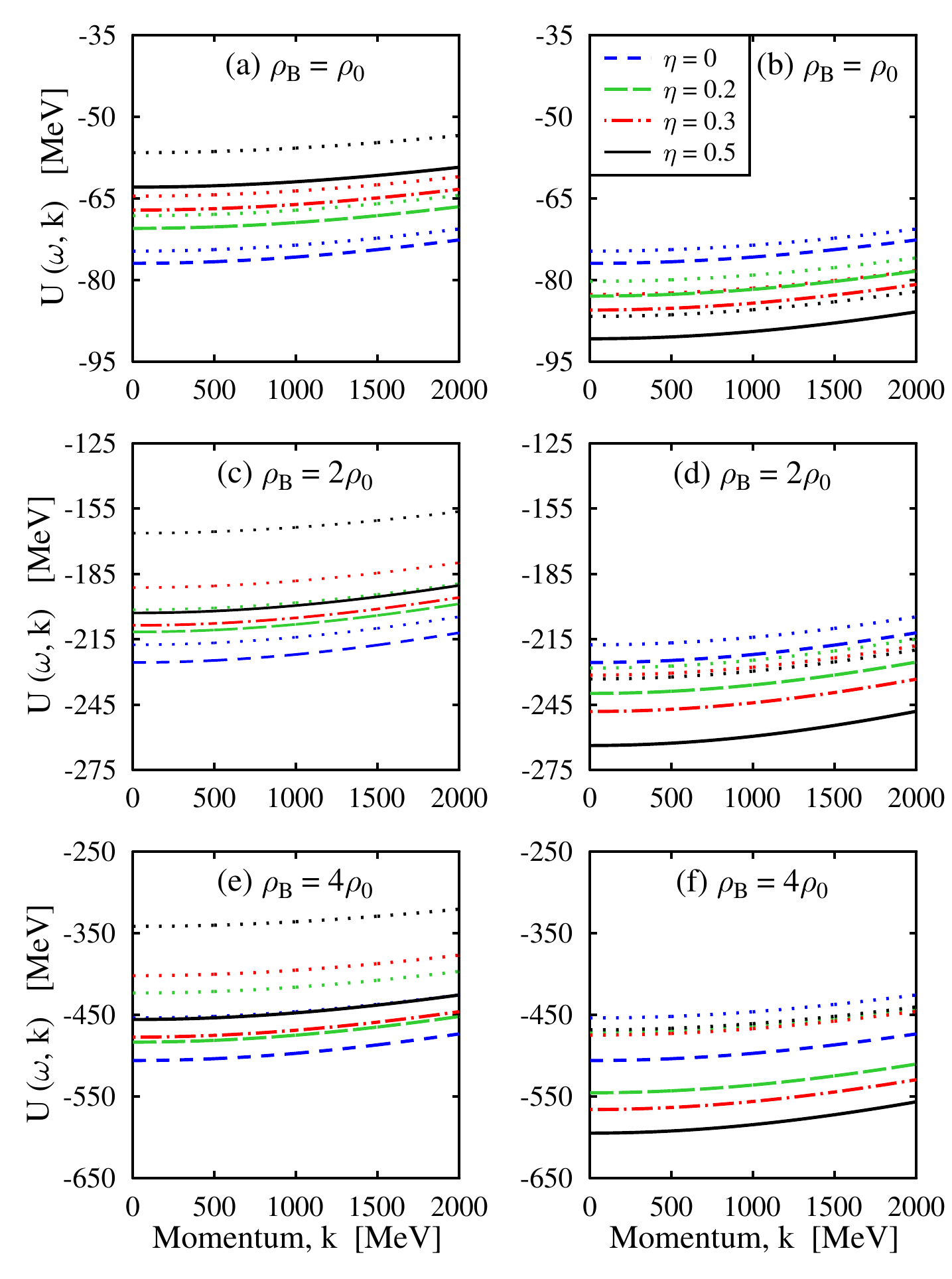}}\caption{ \label{OptPot_Bbar_fs0fs3} (Color Online) The optical potentials of the \Bbarm s (\bminus \ in subplots (a), (c) and (e), and \bbarzero \ in subplots (b), (d) and (f)), as a function of momentum $k \ (\equiv |{\vec k}|)$,  in cold $(T=0)$ hyperonic matter (with $f_s = 0.3$), at various values of the \isoap \ ($\eta$), as described in the legend, and at different densities. In each case, this hyperonic matter situation is also compared against the nuclear matter ($f_s = 0$) situation, represented by dotted lines. }  
\end{center}
\end{figure}

\begin{figure}
\begin{center}
\scalebox{0.59}{\includegraphics{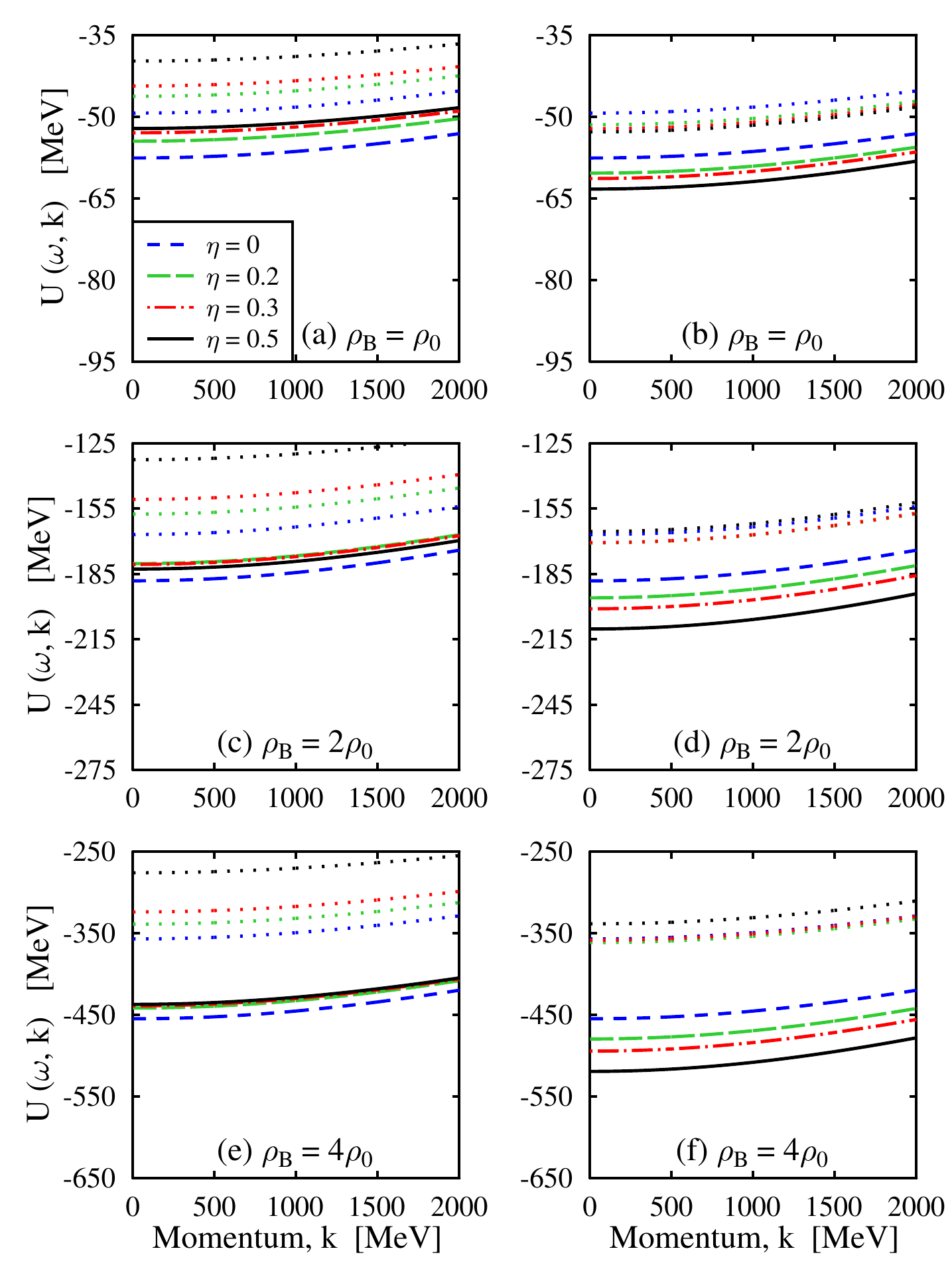}}\caption{ \label{OptPot_B_fs0fs5} (Color Online) The optical potentials of the \Bm s (\bplus \ in subplots (a), (c) and (e), and \bzero \ in subplots (b), (d) and (f)), as a function of momentum $k \ (\equiv |{\vec k}|)$,  in cold $(T=0)$ hyperonic matter (with $f_s = 0.5$), at various values of the \isoap \ ($\eta$), as described in the legend, and at different densities. In each case, this hyperonic matter situation is also compared against the nuclear matter ($f_s = 0$) situation, represented by dotted lines.}  
\end{center}
\end{figure}
\begin{figure}
\begin{center}
\scalebox{0.59}{\includegraphics{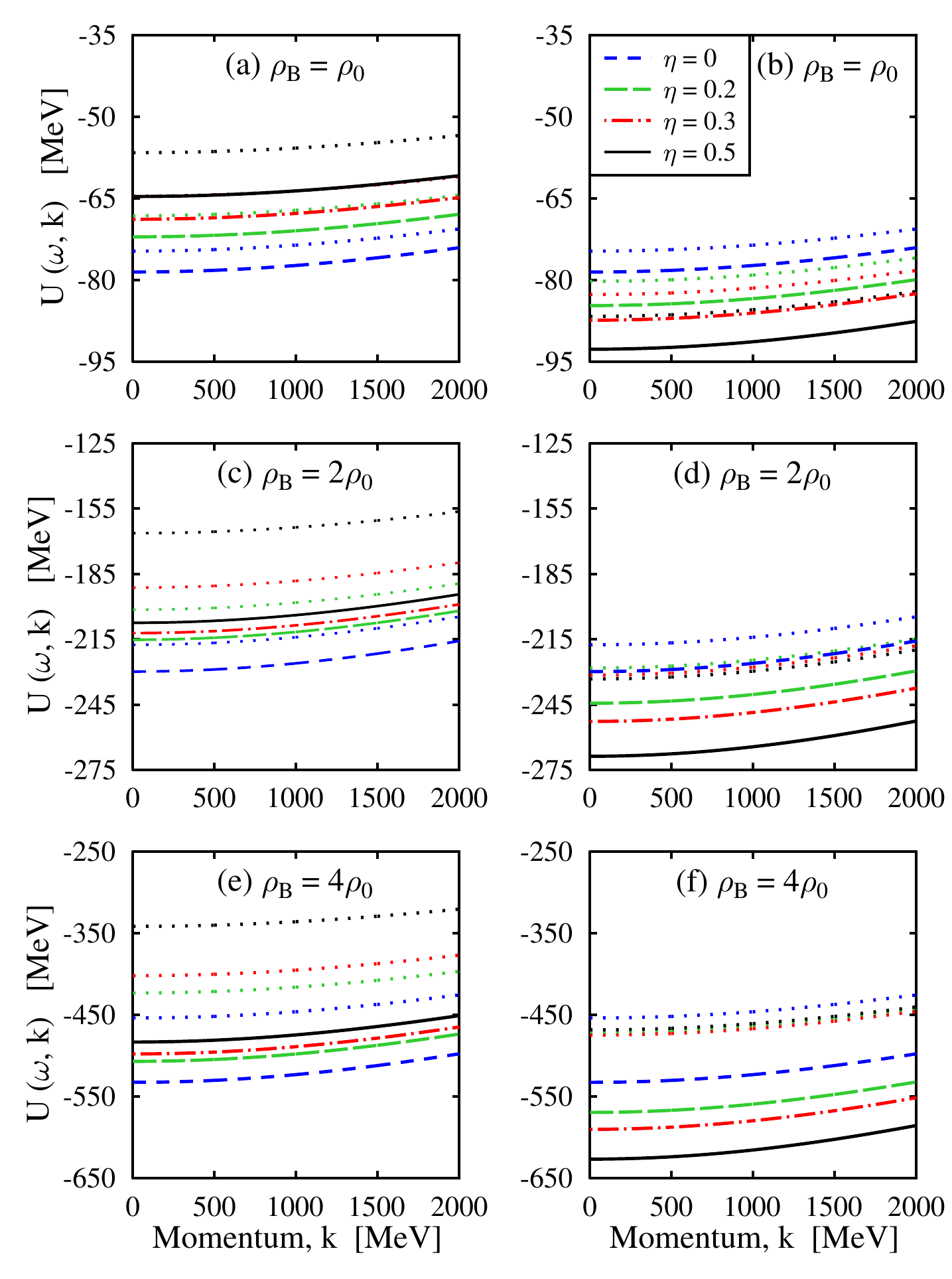}}\caption{ \label{OptPot_Bbar_fs0fs5} (Color Online) The optical potentials of the \Bbarm s (\bminus \ in subplots (a), (c) and (e), and \bbarzero \ in subplots (b), (d) and (f)), as a function of momentum $k \ (\equiv |{\vec k}|)$,  in cold $(T=0)$ hyperonic matter (with $f_s = 0.5$), at various values of the \isoap \ ($\eta$), as described in the legend, and at different 
densities. In each case, this hyperonic matter situation is also compared against the nuclear matter ($f_s = 0$) situation, represented by dotted lines. }  
\end{center}
\end{figure}

Finally, we compare the results of our investigation with the existing treatments of the \BandBbarm \ in-medium properties, using approaches other than this chiral effective model. 
In the quark-meson coupling approach of Ref.\cite{Bm_QMC_Tsushima}, the in-medium mass of various pseudo-scalar and vector mesons, as well as for baryons, was studied as a function of total baryonic density of the medium. The \Bm s are observed to undergo, at the nuclear saturation density for example, a mass drop of about $ 60 \ {\rm MeV}$ from its vacuum value, which is in good agreement with the $49$ MeV drop that follows from our analysis. However, we notice that this approach does not distinguish between the \BandBbarm s at all, which implies that it would be more sensible to compare this number against the average mass drop of \BandBbarm s from our analysis, which stands at $61.5$ MeV (average of $49$ MeV mass drop for the \Bm s and the $74$ MeV drop for the \Bbarm s). Thus, both, the attractive nature of the interactions, as well as the magnitude of the mass drop, are in good agreement with what follows from this generalized chiral effective approach. 
An attractive nature of the in-medium interaction was also observed in the approach of Ref.\cite{YasuiSudoh_Bm_2013_1}, where the \Bm \ $-$ nucleon interaction 
was considered to take place exclusively through pion exchange.
Here, the \Bm s were found to undergo a mass drop of $106$ MeV in isospin symmetric nuclear matter, at $\grb = \grz$. We note however, that in this work, 
the authors have used $\grz = 0.17 \ {\rm fm}^{-3}$, in contrast with the
 $\grz = 0.15 \ {\rm fm}^{-3}$ used in both this investigation, as well as 
in the QMC approach of Ref.\cite{Bm_QMC_Tsushima}.
 At $\grb = 0.17 \ {\rm fm}^{-3}$, 
the value of mass drop for the \Bm s in 
isospin symmetric nuclear matter in our work stands at $63$ MeV, 
which is still, appreciably smaller as compared to their study.  
We also point out that the mass degeneracy of \bplus \ and \bzero \ mesons 
in isospin symmetric matter, and a mass splitting between these isospin 
pairs in the asymmetric situation, which we had observed 
earlier in this section, is exactly replicated in the treatment of 
Ref.\cite{YasuiSudoh_Bm_2013_1}. Thus, there is a qualitative 
agreement between the results of these two approaches. The same authors 
have also adopted a different approach towards analyzing the in-medium 
behavior of \Bm s in Ref.\cite{YasuiSudoh_Bm_2014}, where these in-medium 
interactions have been considered from the point of view of heavy meson 
effective theory, with $1/M$ corrections. This approach provides 
corresponding mass drops of $42$ and $32$ MeV corresponding to two
 different sets of parameter choices, at $\grb = 0.17 \ {\rm fm}^{-3}$, 
which are in even closer agreement with what we have found in this 
investigation, than the approach of Ref.\cite{YasuiSudoh_Bm_2013_1}. 
A similar attractive nature of the $B-N$ interaction also follows 
from the analysis of Ref.\cite{YasuiSudoh_BN_bindingenergy2009}, 
where the $J_P = (1/2)^- \ BN$ state was reported to have a binding 
energy of $9.4$ MeV, hence implying a stable bound state. 
The masses of the \Bm s were also observed to drop in calculations employing the QCD Sum Rules approach \cite{Hilger_QSR_D_Bm, Bm_QSR_arxiv_may2014}; however, an increase in the medium mass for the \Bbarm s was observed in \cite{Hilger_QSR_D_Bm}, which is in contrast to the findings of this investigation.
In the present work, the masses and optical potentials of the $B$ and $\bar B$ mesons in hadronic matter have been studied using an effective chiral model generalized to include 
the bottom sector, to derive the interactions of these mesons 
to the light hadrons. A systematic analysis of the effects of 
density and temperature,
as well as sensitivity of the in-medium properties of the $B$ and
$\bar B$ mesons to 
isospin asymmetry 
and strangeness fraction of the hadronic medium have been carried out in the present investigation.   
\end{section}

\begin{section}{Summary} 
To summarize, we have studied the in-medium masses of the \BandBbarm s in hot 
and dense strange hadronic medium. To this end, we consider a generalization of a chiral effective model originally designed for the light quark sector. However, due to the large mass of $b$ quark, it stays frozen in the medium, and all medium modifications are due to the light quark (or antiquark) content of these mesons.  
Progressively building from (isospin) symmetric cold nuclear matter to symmetric cold hyperonic matter, to include the effects of finite temperatures, to further venture into the territory of asymmetric matter, we have systematically studied the dependence of the in-medium mass of these \BandBbarm s on baryonic density, temperature, strangeness and isospin asymmetry in the medium. We find that each of these mesons experiences a net attractive interaction in the medium, and possesses an in-medium mass 
smaller than its vacuum value 
at all finite densities. These medium effects are found to be strongly 
density dependent, with the medium mass progressively decreasing as 
we go to higher densities. We have restricted our discussions of
the density effects on the masses $B$ and $\bar B$ mesons to about
4 times nuclear matter density. This is because, at still higher densities,
the chiral effective model loses its applicability, when the hadrons 
are no longer the degrees of freedom, as the system undergoes 
a transition to quark matter.
In the present investigation, we find the medium effects to be weakly 
temperature dependent, this weak dependence extending over the entire 
regime in which a hadronic phase is believed to exist (and hence, 
this chiral effective approach can apply). We find that the effect 
of addition of strangeness to the medium is to intensify the mass 
drops for both \BandBbarm s, 
hence implying that the medium becomes more attractive with the addition of hyperons. The vectorial \wtt \ has equal and opposite contributions for the \Bm s and \Bbarm s, resulting in the fact that they have unequal medium masses; however, the two \Bm s (as well as the two \Bbarm s) are degenerate in isospin symmetric matter. In an isospin asymmetric medium, however, even this degeneracy gets broken, and all four of these mesons possess unequal masses, with the \bbarzero \ meson experiencing the largest amount of mass drop in the medium. Also, 
for all of them, 
this dependence on density, asymmetry and strangeness is also reflected in their in-medium optical potentials. Each of these observed features finds a requisite explanation from the point of view their self energies in the medium, derived from their interaction \lagd \ in 
this chiral effective model. 
The in-medium behavior we find on the basis of this generalization of this chiral effective model, bodes well with independent calculations based on alternative methods, wherever such a comparison is possible. To the best of our knowledge, the effect of strangeness and temperature on the in-medium properties of \BandBbarm s, as well as an analysis of the in-medium behavior of these mesons at densities larger than the normal matter density (\grz), are all features hitherto unconsidered in the literature.    
\end{section}  

\begin{section}*{Acknowledgments} 
A.M. would like to thank Department of Science and Technology, Government of India (Project No. SR/S2/HEP-031/2010) for financial support. D.P. acknowledges financial support from University Grants Commission, India [Sr. No. 2121051124, Ref. No. 19-12/2010(i)EU-IV].

\end{section}



\begin{thebibliography}{999}

\bibitem{Rev_1997}
C. M. Ko, V. Koch and G. Q. Li, Annu. Rev. Nucl. Part. Sci. {\bf 47}, 505 (1997).

\bibitem{RMP2010}
R. S. Hayano and T. Hatsuda, Rev. Mod. Phys. {\bf 80}, 2949 (2010).

\bibitem{Li_Rev_99}
G. Q. Li, Prog. Part. Nucl. Phys. {\bf 43}, 619 (1999).

\bibitem{glen}
N. K. Glendenning, \emph{Compact Stars $-$ Nuclear Physics, Particle Physics and General Relativity}, $2^{\rm nd}$ edition. (Springer, 2000).

\bibitem{Oset_PRL98}
J. A. Oller, E. Oset and J. R. Pel\'aez, \PRLett{80}{3452}{1998}

\bibitem{Oset_NPA98}
E. Oset and A. Ramos, \NPA{635}{99}{1998}

\bibitem{mam_cpldchnl_2004}
L. Tol\'os, J. Schaffner-Bielich and A. Mishra, \PRC{70}{025203}{2004}

\bibitem{hofmannlutz}
J. Hofmann and M. F. M. Lutz, \NPA{763}{90}{2005}

\bibitem{QSR_original}
M. A. Shifman, A. I. Vainshtein and V. I. Zakharov, \NPB{147}{448}{1979}

\bibitem{hayashigaki}
A. Hayashigaki, \PLB{487}{96}{2000} 

\bibitem{Hilger_QSR_D_Bm}
T. Hilger, R. Thomas and B. K{\"a}mpfer, \PRC{79}{025202}{2009}

\bibitem{arvQSR}
A. Kumar and A. Mishra, \PRC{82}{045207}{2010}

\bibitem{QMC1}
K. Tsushima, D. H. Lu, A. W. Thomas, K. Saito and R. H. Landau, \PRC{59}{2824}{1999}

\bibitem{QMC2}
A. Sibirtsev, K. Tsushima and A. W. Thomas, \EPJA{6}{351}{1999}

\bibitem{Bm_QMC_Tsushima}
K. Tsushima and F. C. Khanna, \PLB{552}{138}{2003}

\bibitem{serotwalecka}
B. D. Serot and J. D. Walecka, Int. J. Mod. Phys. E {\bf6}, 515 (1997).

\bibitem{kaplan_nelson}
D. B. Kaplan and A. E. Nelson, \PLB{175}{57}{1986}

\bibitem{li_ko_brown_prl95}
G. Q. Li, C. M. Ko and G. E. Brown, \PRLett{75}{4007}{1995}

\bibitem{li_ko}
G. Q. Li and C. M. Ko, \NPA{582}{731}{1995}

\bibitem{li_ko_brown}
G. Q. Li, C. M. Ko and G. E. Brown, \NPA{606}{568}{1996}

\bibitem{Jpsi_suppression_decays_mesons}
A. Sibirtsev, K. Tsushima, K. Saito and A. W. Thomas, \PLB{484}{23}{2000} 

\bibitem{ParticleRatiosJPhysG_1}
Q. Li, Z. Li, S. Soff, M. Bleicher and H. St\"ocker, \JPhysG{32}{407}{2006}

\bibitem{Braun_Munzinger_Production_Asymmetry}
A. Andronic, P. Braun-Munzinger, K. Redlich and J. Stachel, \PLB{659}{149}{2008}


\bibitem{Pap_prc99}
P. Papazoglou, D. Zschiesche, S. Schramm, J. Schaffner-Bielich, H. St\"ocker and W. Greiner, \PRC{59}{411}{1999}

\bibitem{Zsch}
D. Zschiesche et al., \emph{`Chiral Symmetries in Nuclear Physics'}, in \emph{Symmetries in Intermediate and High Energy Physics}, Springer Tracts in Modern Physics, Vol. 163 (Springer-Verlag, 2000).

\bibitem{mamZsch_vector2004}
D. Zschiesche, A. Mishra, S. Schramm, H. St\"ocker and W. Greiner, \PRC{70}{045202}{2004}

\bibitem{mam_balazs_vector2004}
A. Mishra, K. Balazs, D. Zschiesche, S. Schramm, H. St\"ocker and W. Greiner, \PRC{69}{024903}{2004}

\bibitem{mamK2004} 
A. Mishra, E. L. Bratkovskaya, J. Schaffner-Bielich, S. Schramm and H. St\"ocker, \PRC{70}{044904}{2004}

\bibitem{mam_kaons2006}
A. Mishra and S. Schramm, \PRC{74}{064904}{2006}

\bibitem{mam_kaons2008}
A. Mishra, S. Schramm and W. Greiner, \PRC{78}{024901}{2008}

\bibitem{sambuddha1}
A. Mishra, A. Kumar, S. Sanyal and S. Schramm, \EPJA{41}{205}{2009}

\bibitem{sambuddha2}
A. Mishra, A. Kumar, S. Sanyal, V. Dexheimer and S. Schramm, \EPJA{45}{169}{2010}

\bibitem{mamD2004} 
A. Mishra, E. L. Bratkovskaya, J. Schaffner-Bielich, S. Schramm and H. St\"ocker, \PRC{69}{015202}{2004} 

\bibitem{arindam}
A. Mishra and A. Mazumdar, \PRC{79}{024908}{2009} 

\bibitem{arvDprc}
A. Kumar and A. Mishra, \PRC{81}{065204}{2010}

\bibitem{arvDepja}
A. Kumar and A. Mishra, \EPJA{47}{164}{2011}

\bibitem{leeko} 
S. H. Lee and C. M. Ko, \PRC{67}{038202}{2003}

\bibitem{leeko2} 
S. H. Lee and C. M. Ko, Prog. Theor. Phys. Suppl. {\bf 149}, 173 (2003).

\bibitem{DP_Bmonia}
A. Mishra and D. Pathak, Phys. Rev. C {\bf 90}, 025201 (2014).

\bibitem{Tolos_Bm}
J. M. Torres-Rincon, L. Tolos and O. Romanets, \PRD{89}{074042}{2014}

\bibitem{MooreTeaney_HeavyQuarkThermalization}
G. D. Moore and D. Teaney, \PRC{71}{064904}{2005}

\bibitem{Fries_heavyQ_PRC2012}
M. He, R. J. Fries and R. Rapp, \PRC{86}{014903}{2012}

\bibitem{Bm_Transport_MesonGas}
L. M. Abreu, D. Cabrera and J. M. Torres-Rincon, \PRD{87}{034019}{2013}

\bibitem{JaneAlam_Bm}
S. K. Das, S. Ghosh, S. Sarkar and J. Alam, \PRD{85}{074017}{2012}

\bibitem{HeFriesPLB_May2014}
M. He, R. J. Fries and R. Rapp, \PLB{735}{445}{2014}

\bibitem{BaBar_Offical_Webpage}
BaBar Collaboration, Official Webpage, http://www-public.slac.stanford.edu/babar

\bibitem{BELLE_Official_Webpage}
BELLE Collaboration, Official Webpage, http://belle.kek.jp

\bibitem{BELLE_PTEP2012}
J. Brodzicka et al. (BELLE Collaboration), Prog. Theor. Exp. Phys. 04D001 (2012).

\bibitem{BELLE2_Official_Webpage}
BELLE-II Collaboration, Official Webpage, http://belle2.kek.jp

\bibitem{BELLE2_DESY}
BELLE-II experiment, Official Website of DESY, http://belle2.desy.de

\bibitem{BELLE2_Technical_Design_Report}
BELLE-II Technical Design Report, KEK Report 2010-1, arXiv:1011.0352 [physics.ins-det]

\bibitem{YasuiSudoh_Bm_2014}
S. Yasui and K. Sudoh, \PRC{89}{015201}{2014}

\bibitem{YasuiSudoh_Bm_2013_1}
S. Yasui and K. Sudoh, \PRC{87}{015202}{2013}

\bibitem{YasuiSudoh_Bm_2013_2}
S. Yasui and K. Sudoh, \PRC{88}{015201}{2013}

\bibitem{weinberg67}
S. Weinberg, \PRLett{18}{188}{1967}

\bibitem{weinberg68}
S. Weinberg, Phys. Rev. {\bf 166}, 1568 (1968).

\bibitem{coleman1}
S. Coleman, J. Wess and B. Zumino, Phys. Rev. {\bf 177}, 2239 (1969).

\bibitem{coleman2}
C. G. Callan, S. Coleman, J. Wess and B. Zumino, Phys. Rev. {\bf 177}, 2247 (1969).

\bibitem{bardeenlee}
W. A. Bardeen and B. W. Lee, Phys. Rev. {\bf 177}, 2389 (1969).

\bibitem{brownrho_PRL1991}
G. E. Brown and M. Rho, \PRLett{66}{2720}{1991}

\bibitem{ellis}

E. K. Heide, S. Rudaz, P. J. Ellis, Nucl. Phys. A {\bf 571}, 713 (2001).

\bibitem{schechter}

J. Schechter, Phys. Rev. D {\bf 21} 3393 (1980).

\bibitem{klinglkimlee}

F. Klingl, S. Kim, S. H. Lee, P. Morath, and W. Weise, 
Phys. Rev. Lett. {\bf 82}, 3396 (1999).

\bibitem{PDG2012}
J. Beringer et al. (Particle Data Group), Phys. Rev. D {\bf 86}, 010001 (2012), and 2013 partial update for the 2014 edition.

\bibitem{LQCD1_fB}
A. Bazavov et al. (Fermilab Lattice and MILC Collaboration), \PRD{85}{114506}{2012}  

\bibitem{LQCD2_fB}
H. Na et al. (HPQCD Collaboration), \PRD{86}{034506}{2012}

\bibitem{QSR1_fB}
S. Narison, \PLB{718}{1321}{2013}

\bibitem{QSR2_fB}
S. Narison, \PLB{721}{269}{2013}

\bibitem{QSR3_fB}
P. Gelhausen, A. Khodjamirian, A. A. Pivovarov and D. Rosenthal, \PRD{88}{014015}{2013}

\bibitem{ETM_fB}
M. J. Baker et al. (ETM Collaboration), JHEP07, 032 (2014).

\bibitem{fB_Lucha}
W. Lucha, D. Melikhov, and S. Simula, J. Phys. G {\bf 38}, 105002 (2011).

\bibitem{LQCD3_fB}
R. J. Dowdall, C. T. H. Davies, R. R. Horgan, C. J. Monahan, 
and J. Shigemitsu (HPQCD Collaboration), \PRLett{110}{222003}{2013}

\bibitem{roder}
D. R\"oder, J. Ruppert and D. H. Rischke, \PRD{68}{016003}{2003}

\bibitem{LQCD1}
Y. Aoki, G. Endr\H{o}di, Z. Fodor, S. D. Katz and K. K. Szab\'{o}, Nature {\bf443}, 675 (2006).

\bibitem{LQCD2}
C. Bernard et al. (MILC Collaboration), \PRD{71}{034504}{2005}

\bibitem{LQCD3}
A. Bazavov et al. (HotQCD Collaboration), \PRD{85}{054503}{2012};
A. Bazavov et al., Phys. Rev. Lett. {\bf 113}, 072001 (2014).

\bibitem{leshouches}
W. Weise - \emph{``Quarks, Hadrons and Dense Nuclear Matter''}, lectures
in proceedings of “Trends in Nuclear Physics, 100 years later”, Les Houches, 
Session LXVI, 1996.

\bibitem{YasuiSudoh_BN_bindingenergy2009}
S. Yasui and K. Sudoh, \PRD{80}{034008}{2009}

\bibitem{Bm_QSR_arxiv_may2014}
K. Azizi, N. Er and H. Sundu, arXiv:1405.3058 [hep-ph]

\end{thebibliography}
\end{document}